\documentclass[journal,12pt,onecolumn,draftclsnofoot,]{IEEEtran}
\usepackage{xmpaper,lzqsymbol}
\usepackage{amssymb,amsmath,latexsym}
\usepackage{graphicx,subfigure}
\usepackage{color,cite,times}
\usepackage{epstopdf}
\usepackage{multirow}
\usepackage{array}

\usepackage{amsmath}
\usepackage{footnote}
\usepackage{bm}
\usepackage{algorithm}
\usepackage{algorithmic}
\usepackage{bm}
\usepackage[T1]{fontenc}

\usepackage{booktabs}
\usepackage{bbm} 
\usepackage{subfigure}

\begin{document}
\title{Scattering Environment Aware Joint Multi-user Channel Estimation and Localization with Spatially Reused Pilots}
\DeclareRobustCommand*{\IEEEauthorrefmark}[1]{%
    \raisebox{0pt}[0pt][0pt]{\textsuperscript{\footnotesize\ensuremath{#1}}}}

\author{Kaiyuan Tian, Yani Chi, Yufan Zhou, An Liu,~\IEEEmembership{Senior~Member, IEEE}
\thanks{Kaiyuan Tian is with the  State Key Laboratory of Advanced Optical Communication Systems and Networks,  School of Electronics, Peking University, Beijing 100871, China. (e-mail: tiankaiyuan00@gmail.com). 
 
Yani Chi is with the China Academy of Information and Communications Technology, Beijing 100191, China.(e-mail: chiyani@caict.ac.cn). 

Yufan Zhou and An Liu are with the College of Information Science and Electronic Engineering, Zhejiang University, Hangzhou
310027, China (email: yufanzhou@zju.edu.cn, anliu@zju.edu.cn).
}
}

\maketitle

\begin{abstract}
The increasing number of users leads to an increase in pilot overhead, and the limited pilot resources make it challenging to support all users using orthogonal pilots. By fully capturing the inherent physical characteristics of the multi-user (MU) environment, it is possible to reduce pilot costs and improve the channel estimation performance.  
In reality, users nearby may share the same scatterer, while users further apart tend to have orthogonal channels. This paper proposes a two-timescale approach for joint MU uplink channel estimation and localization in MIMO-OFDM systems, which fully captures the spatial characteristics of MUs. To accurately represent the structure of the MU channel, the channel is modeled in the 3-D location domain. In the long-timescale phase, the time-space-time multiple signal classification (TST-MUSIC) algorithm initially offers a rough approximation of scatterer positions for each user, which is subsequently refined through the scatterer association algorithm based on density-based spatial clustering of applications with noise (DBSCAN) algorithm. The BS then utilizes this prior information to apply a graph-coloring-based user grouping algorithm, enabling spatial division multiplexing of pilots and reducing pilot overhead. In the short timescale phase, a low-complexity scattering environment aware location-domain turbo channel estimation (SEA-LD-TurboCE) algorithm is introduced to merge the overlapping scatterer information from MUs, facilitating high-precision joint MU channel estimation and localization under spatially reused pilots.  Simulation results verify the superior channel estimation and localization performance of our proposed scheme over the baselines.

\end{abstract}

   \section{Introduction}

Massive multiple input multiple output (MIMO) technology is crucial for the development of 5G and future generations of wireless systems. It utilizes both spatial multiplexing and array gains to meet the growing demand for high-speed data transmission~\cite{larsson2014massive},~\cite{rusek2012scaling}. At the same time, orthogonal frequency division multiplexing (OFDM) has gained recognition as an effective modulation technique for wireless communications. It is well regarded for its ability to provide reliable high-speed data transfer and its resilience against frequency selectivity~\cite{cimini1985analysis}. Consequently, the combination of massive MIMO and OFDM has become a vital physical layer technology in the framework of 5G new radio (NR) standards~\cite{lu2014overview}.

A critical challenge in the deployment of massive MIMO-OFDM systems revolves around the accurate acquisition of channel state information (CSI) at the base station (BS) within the constraints of limited pilot resources. The time division duplex (TDD) systems offer a significant advantage over frequency division duplex (FDD) counterparts by markedly reducing the pilot overhead necessary for CSI acquisition at the BS~\cite{xia2013}. In TDD systems, the BS estimates the uplink channel utilizing the received sounding reference signals (SRSs) as stipulated in 5G NR~\cite{3gpp2021}. Subsequently, this estimated uplink channel informs the downlink precoding, capitalizing on the channel reciprocity characteristic of TDD systems~\cite{chien2017}. 

The current 5G communication framework employs orthogonal uplink pilots (such as code domain multiplexing, time domain multiplexing, and frequency domain multiplexing) to estimate MU channels, thus preempting pilot-to-pilot interference \cite{ref33}. However, as the mobile user population continues to swell, the finite number of orthogonal pilots in forthcoming 5.5G large-scale antenna communication systems may not meet the requirements to effectively serve the entire user base\cite{ref34}. This impending limitation calls for innovative solutions to optimize pilot utilization and enhance the overall efficiency of channel estimation in the face of increasing user density. Besides, the pilots can also provide information about the positions of users and scatterers. Thus, the problem of joint MU channel estimation and scatterer/user localization by pilots is a promising area, especially in future multi-functional wireless systems that also need to provide high-accuracy localization and sensing performance. We summarize some related works below.

\textbf {MU pilots design and assignment:}
In MU systems, orthogonal resources such as time domain, frequency domain, and code domain can be utilized to allocate pilots. However, the use of orthogonal pilots limits the number of users that the system can accommodate. Non-orthogonal pilot transmission can increase the number of users the system can support, but it necessitates additional processing to manage inter-user interference. Current non-orthogonal pilot assignment principles primarily focus on power-based considerations. For instance, the graph-coloring-based algorithms are employed on the same time-frequency resources, allowing users with strong interference to adopt different pilot sequences to reduce interference~\cite{ref7}. In~\cite{ref8}, a pilot power assignment policy was developed to maximize the minimum asymptotic signal-to-noise plus interference ratio (SINR) in each cell, by adopting a pilot scheme where all users in each cell share the same pilot sequence and keep the pilot sequences orthogonal for different cells. In~\cite{ref9}, diversified user channels are recognized, with a larger number of pilots allocated to users with typical channels and fewer pilots assigned to users in extreme environments. Additionally, research on channel estimation using non-orthogonal Zadoff-Chu (ZC) sequences has been conducted~\cite{ref10}. To address the inter-user interference, an algorithm based on the multi-stage successive interference cancellation (MSIC) technique and the Turbo algorithm framework has been proposed for MU channel estimation.

\textbf {MU joint channel estimation:}
Research in MU massive MIMO systems has identified that channels from physically proximate users can exhibit shared sparse structures due to common local scattering clusters~\cite{ref1},~\cite{ref2}. This shared sparsity can enhance sparse signal recovery, improving downlink channel estimation performance. In TDD systems, a prior-based hidden Markov model~\cite{ref3} has been proposed for calculating transmitters' downlink CSI, using pilot data from various users for joint angle domain channel estimation at the BS. As for FDD systems, a robust closed-loop pilot and CSI feedback resource adaptation framework~\cite{ref4} leverage joint sparsity in the angle domain to enhance CSI estimation. A joint orthogonal matching pursuit (OMP) recovery algorithm~\cite{ref5} capitalizes on common sparsity between users, though its success requires that all users share a large portion of common channel support. A general model~\cite{ref6} incorporating both shared and individual sparsity components has been suggested for massive MIMO downlink channel estimation. However, these methods' assumption that closely located users' channels share the common angular support is an oversimplification. In fact, nearby users may share scatterers, leading to common support in the 3-D location domain.

\textbf {Joint scatterer/user localization and channel estimation}: In \cite{zhou2021simultaneous}, a massive MIMO-OFDM channel was modeled based on the position of scatterers and a user, and then the user location and channel coefficients were simultaneously estimated. In \cite{zheng2020joint}, a dynamic grid-based method was proposed to improve user localization and channel estimation performance. In \cite{{yang2022soft}}, the authors proposed to provide soft information about channel estimation and user location instead of hard information about those. In \cite{hong2022joint}, two geometry-based models were proposed for performing joint channel estimation and scatterer localization involved in different propagation paths of bouncing order. The authors in \cite{ref32} studied the joint location sensing and channel estimation problem of an intelligent reflecting surface (IRS) aided millimeter wave system, which takes advantage of the partially overlapping structured sparsity to enhance the overall performance.

However, the existing channel estimation and localization algorithms for multi-user (MU) environments fail to fully harness the spatial information available. Specifically, two key aspects remain unexplored: 1) The overlapping information of MU channels in the 3-D location domain, which will be elaborated in Section \ref{sec:system_model}, has not been utilized. 2) The potential of spatial division multiplexing for pilot allocation in MU scenarios has yet to be fully realized. Additionally, the challenge of jointly performing MU channel estimation and localization of scatterers/users under conditions of spatially reused uplink pilots requires further investigation.

Motivated by the limitations of current methods and the constraints imposed by limited pilot resources, this paper introduces a novel two-timescale joint MU channel estimation and localization scheme for TDD massive MIMO-OFDM systems. The proposed approach aims to effectively capitalize on the spatial characteristics inherent in MU environments. To achieve this, we leverage the overlapping or orthogonal channel properties unique to different MUs. The primary contributions of this paper are as follows:


\textbf{(1) Two-timescale Joint MU Channel Estimation and Localization Scheme:} This paper introduces a novel two-timescale framework for joint MU channel estimation and localization. In the long timescale phase, the time-space-time multiple signal classification (TST-MUSIC) algorithm is employed to obtain a coarse estimation of scatterer positions for each user. To enhance the accuracy of these estimates, we integrate the overlapping scatterer information across multiple users by applying the density-based spatial clustering of applications with noise (DBSCAN) algorithm. Leveraging this refined scatterer information, the BS implements a graph-coloring-based user grouping strategy. This approach enables efficient spatial division multiplexing of pilots among users within the same group, thereby minimizing pilot overhead. In the short timescale phase, we propose a low-complexity scattering environment-aware location-domain turbo channel estimation (SEA-LD-TurboCE) algorithm. This algorithm facilitates high-precision joint MU channel estimation and localization under spatially reused pilots.

\textbf {(2) Graph-coloring-based user grouping algorithm for spatial pilot reuse:}
Multiple multiplexing methods exist for achieving MU orthogonal pilot transmission, including code-domain multiplexing, time-domain multiplexing, and frequency-domain multiplexing. To enhance user capacity, location-domain multiplexing of pilots is introduced, allowing MUs to transmit the same pilots simultaneously on the same frequency.  However, the key to this expansion is to eliminate spatial interference among users sharing the same pilot, necessitating a grouping strategy that minimizes intra-group interference.  The minimum distance between MU scatterers serves as a measure for grouping, with the interference relationships among users graphically depicted.  This transforms the issue into a graph coloring problem and can be solved by the particle swarm optimization (PSO) algorithm.  With users in the same group being spatially near-orthogonal, the SEA-LD-TurboCE algorithm can then jointly estimate channels for MUs and automatically mitigate interference.

\textbf {(3) SEA-LD-TurboCE algorithm under spatially reused pilots:}
For the short timescale phase, we propose a low-complexity message passing algorithm to address the MU joint channel estimation and localization problem under spatially reused pilots. This algorithm leverages the prior information of scatterer positions acquired during the long-timescale phase to initialize a dynamic grid in the location domain. This initialization not only bolsters estimation accuracy but also reduces computational complexity. Furthermore, the SEA-LD-TurboCE algorithm can distinguish the channels of different users even when non-orthogonal pilots are employed, which is facilitated by the location domain modeling and proper assignment of pilots.




The rest of the paper is organized as follows. In Section \ref{sec:system_model}, we present the system model. In Section \ref{sec:Channel Estimation Scheme}, we introduce the two-timescale joint MU channel estimation and localization scheme in the location domain channels. 
In Section \ref{sec:Long-Timescale Algorithm}, we introduce the long-timescale algorithm for scatterer localization and pilot assignment. In Section \ref{sec:DBSCAN-Turbo-SBI Algorithm}, we present the proposed SEA-LD-TurboCE algorithm under spatially reused pilots. Simulation results and conclusion are given in Section \ref{sec:simulation_results} and \ref{sec:conclusion}, respectively.

    \section{System Model }\label{sec:system_model}
    \subsection{Channel Model}
    We consider a TDD massive MIMO-OFDM system with one BS with $N=N_x\times N_y$ uniform planar array (UPA) antennas and $K$ users indexed by $k \in \{1, ..., K\}\triangleq  \mathcal{K}$ with a single antenna.  In the spherical coordinate system, the BS's array is located at $\boldsymbol{p}_{\mathrm{BS}}=\left[0, 0, 0\right]$,  The user $k$ is located at $\boldsymbol{p}_{\mathrm{U\!E,k}}=\left[\theta_{\mathrm{U\!E,k}}, \phi_{\mathrm{U\!E,k}},r_{\mathrm{U\!E,k}}\right]$ with $\theta_{\mathrm{U\!E,k}}$, $\phi_{\mathrm{U\!E,k}}$ and $r_{\mathrm{U\!E,k}}$ are the azimuth angle, elevation angle and range of user $k$, respectively. We assume that within a 3-D area $\mathcal{R}$, there are a total of $L$ scatterers indexed by $l \in \{1, \ldots, L\} \triangleq \mathcal{L}$. The position of the $l$-th scatterer is denoted by $\boldsymbol{p}_{l} = [\theta_{l}, \phi_{l}, r_{l}]$ for $l \in \mathcal{L}$. Additionally, we assume that there are $L_k$ scatterers contributing to the channel paths between user $k$ and the BS. These scatterers are indexed by the set $\mathcal{L}_k$, which is an arbitrary subset of $\mathcal{L}$ (i.e., $\mathcal{L}_k \subset \mathcal{L}$). Furthermore, we postulate that the BS possesses some prior information about the user's location, derived from existing localization technologies such as GPS.
       \begin{figure}[t]
      \centering
      \includegraphics[width=6.5in]{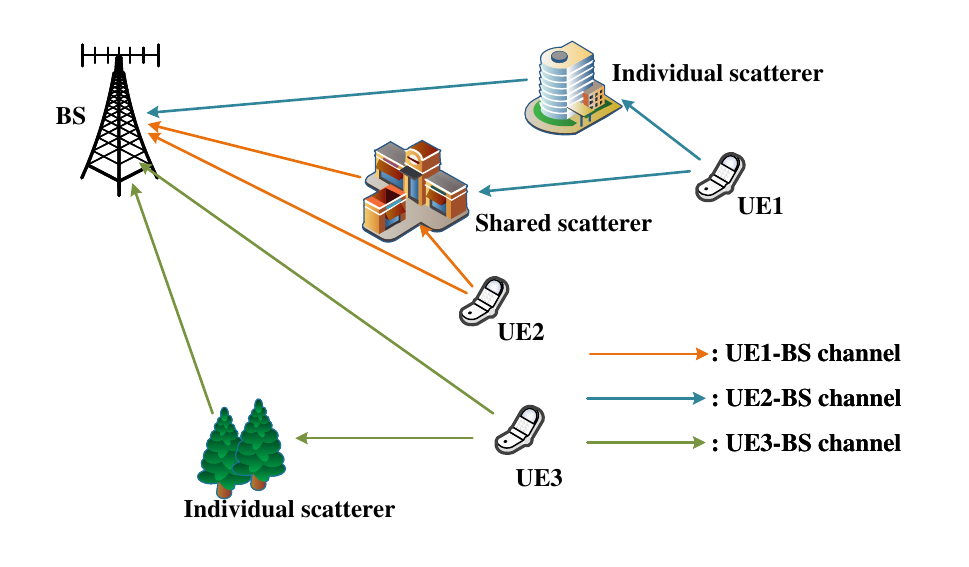}
      \caption{MU uplink channel model.}
      \label{figure_1}
    \end{figure}
On the $p$-th $(1 \leq p \leq P)$ subcarrier, the user $k$ transmits a pilot symbol $u_{k, p}$ and the corresponding received signals related to the BS can be expressed as
\begin{equation}
\boldsymbol{y}_{k, p}=\boldsymbol{h}_{k, p} u_{k, p}+\boldsymbol{z}_{k, p},
\end{equation}
where $\boldsymbol{h}_{k, p} \in \mathbb{C}^{N}$ is the  channel vectors of user $k$ on subcarrier $p$ and $\boldsymbol{z}_{k, p} \sim \mathcal{C N}\left(\mathbf{0},\left(\sigma_{k}\right)^2 \mathbf{I}\right) \in\mathbb{C}^{N}$ is the the additive white Gaussian noise. Note that in (1), we have assumed orthogonal pilots such that there is no interference between different users. In the proposed two-timescale scheme, the orthogonal pilots are used at the long-timescale for localization of scatterers of each user, while spatially reused (non-orthogonal) pilots will be used at the short-timescale for serving more users. The received signal model under the spatially reused pilots will be presented later. For convenience, we define the aggregate received uplink signals of all subcarriers of user $k$ as
$\boldsymbol{y}_k=\left[\left(\boldsymbol{y}_{k, 1}\right)^T, \ldots,\left(\boldsymbol{y}_{k, N}\right)^T\right]^T\in \mathbb{C}^{NP \times 1}$.
The channel vectors can be modeled by the LOS component and NLOS component as follows
\begin{equation}
\boldsymbol{h}_{k, p}=\boldsymbol{h}_{k, p}^{\text{LOS}}+\boldsymbol{h}_{k, p}^{\text{NLOS}},
\end{equation}
with
\begin{equation}
\boldsymbol{h}_{k, p}^{\text{LOS}}=x^k_{0} e^{-j 2 \pi p f_0\tau_{{0}}\left( \boldsymbol{p}_{\mathrm{U\!E},k}\right)}\boldsymbol{a}\left({\theta}_{\mathrm{U\!E},k}, {\phi}_{\mathrm{U\!E},k}\right),
\end{equation}
\begin{equation}
\boldsymbol{h}_{k, p}^{\text{NLOS}}=\sum_{l \in \mathcal{L}_k} x^k_{l} e^{-j 2 \pi p f_0\tau_{{l}}\left( \boldsymbol{p}_{\mathrm{U\!E},k}\right)} \boldsymbol{a}\left(\theta_{l}, \phi_{l}\right),
\end{equation}
where $f_0$ is the subcarrier interval, $c$ is the speed of light, and $\boldsymbol{a}\left(\cdot\right)$ is the array response vector of the UPA antenna array. $x^k_{0}$ and $x^k_{l}$ are the complex gain of the LOS path and the $l$-th NLOS path of user $k$, respectively. Besides, $\tau_{{0}}\left( \boldsymbol{p}_{\mathrm{U\!E},k}\right) = \frac{r_{\mathrm{U\!E,k}}}{c}$ is the delay of the LOS path and $\tau_{{l}}\left( \boldsymbol{p}_{\mathrm{U\!E},k}\right)$ is the delay of the $l$-th NLOS path, which can be expressed as
\begin{equation}
\tau_{{l}}\left( \boldsymbol{p}_{\mathrm{U\!E},k}\right)=\frac{\sqrt{{r_{\mathrm{U\!E},k}^2+r_{l}^2-2 r_{\mathrm{U\!E},k} r_{l}}\left[\sin \left(\theta_{\mathrm{U\!E},k}\right) \sin \left(\theta_{l}\right) \cos \left(\phi_{\mathrm{U\!E},k}-\phi_{l}\right)+\cos \left(\theta_{\mathrm{U\!E},k}\right) \cos \left(\theta_{l}\right)\right]}}{c} +\frac{r_{l}}{c}.
\end{equation}

    \subsection{Location Domain Sparse Representation}

    To fully utilize the overlapping and orthogonal channel characteristics of different MUs, the channel is modeled in the location domain by integrating both user and scatterer positions into the modeling framework. We define a 3-D dynamic location domain grid $\boldsymbol{\gamma}=\left\{\boldsymbol{\gamma}_1, \ldots, \boldsymbol{\gamma}_Q\right\} \subset \mathcal{R}$ of $Q$ location points, where the $q$-th grid point is denoted by $\left(\theta_q, \phi_q, r_q\right)$. Then, we define the sparse basis with the dynamic grid as

\begin{equation} \label{7}
\mathbf{A}(\boldsymbol{\theta}, \boldsymbol{\phi}) \triangleq\left[\boldsymbol{a}\left(\theta_1, \phi_1\right), \ldots, \boldsymbol{a}\left(\theta_Q, \phi_Q\right)\right] \in \mathbb{C}^{N \times Q},
\end{equation}
\begin{equation} \label{8}
\boldsymbol{b}_{k, p}\left(\boldsymbol{\gamma}, \mathbf{p}_{\mathrm{U\!E},k}\right)=\left[\mathrm{e}^{-\mathrm{j} 2 \pi \mathrm{p} f_0\left[\tau\left(\boldsymbol{\gamma}_{1}, \mathbf{p}_{\mathrm{U\!E},k}\right)\right]}, \ldots, \mathrm{e}^{-\mathrm{j} 2 \pi \mathrm{p}f_0\left[\tau\left(\boldsymbol{\gamma}_{ \mathrm{Q}}, \mathbf{p}_{\mathrm{U\!E},k}\right)\right]}\right]^{\mathrm{T}} \in \mathbb{C}^{Q \times 1},
\end{equation}
where $\boldsymbol{\theta} \triangleq\left[\theta_1, \ldots, \theta_Q\right]^T$, $\boldsymbol{\phi} \triangleq$ $\left[\phi_1, \ldots, \phi_Q\right]^T$ and $\boldsymbol{r} \triangleq\left[r_1, \ldots, r_Q\right]^T$. Then, the sparse representation of the NLoS channel vector at $p$-th subcarrier of user $k$ is denoted by
\begin{equation}
\boldsymbol{h}_{k, p}^{\text{NLOS}}=\boldsymbol{A}\left(\boldsymbol{\gamma}\right) \operatorname{Diag}\left(\boldsymbol{b}_{k, p}\left(\boldsymbol{\gamma}, \boldsymbol{p}_{U\!E, k}\right)\right) \boldsymbol{x}_k,
\end{equation}
where $\boldsymbol{x}_k\in \mathbb{C}^{Q \times 1}$ is the location domain sparse NLoS communication channel vector which only has $L_k$ non-zero elements corresponding to $L_k$ scatterers. The $q$-th element of $L_k$, denoted by $x^k_{q}$, represents the complex channel gain of the scatterer lying in the $q$-th gird. Moreover, the non-zero entries of $\boldsymbol{x}_k$ are i.i.d. complex Gaussian distributed with zero mean and unit variance.

    \subsection{MU Joint Sparsity Model}
 Some users, particularly those near one another, might face a very similar scattering structure\cite{ref30}. To exploit the partial overlap scatterers between users, we assume a joint channel sparsity model for MUs. Define $\boldsymbol{s}_k=\left[s^k_1,  \ldots, s^k_{Q}\right]^T$ as the sparsity support of user $k$, where $s_q=1$ if $x^k_{q} \neq 0$ and $s_q=0$ if $x^k_{q} = 0$. Then, we introduce a joint support vector  $\boldsymbol{s} = \left[s_1, \ldots, s_{Q}\right]^T$, where $s_{q} = 0$ if and only if $s^{k}_{q} = 0$ for all users $k$, otherwise $s_{q} = 1$. The MU joint sparsity can be characterized by the conditional probability $p\left(s_q^k \mid s_q\right)$. For example, if the scatterers of two users have a large overlapping, then the support vectors of the two users will share a large number of 1's in the joint support vector, leading to a larger conditional probability $p\left(s_q^k = 1 \mid s_q = 1\right)$.

\section{Two-timescale joint MU channel estimation and localization Problem Formulation}\label{sec:Channel Estimation Scheme}
\subsection{Overall Framework}
Multiplexing methods for MU orthogonal pilot transmission encompass code-domain, time-domain, and frequency-domain multiplexing. Orthogonal pilot transmission effectively eliminates interference during uplink channel estimation among different users. However, the maximum number of supported users is constrained by the number of available orthogonal pilots. As the number of mobile terminals is projected to surge in the near future, this limitation poses a significant bottleneck for current systems. To overcome this limitation and expand pilot resources, we propose the adoption of location-domain multiplexing for pilot signals, aiming to enhance user capacity. The crucial challenge in this expansion lies in mitigating spatial interference among users sharing the same pilot. To address this issue, we devise a two-timescale joint MU channel estimation and localization scheme that can effectively utilize the prior information, as shown in Fig. 2. 
       \begin{figure}[t]
      \centering
      \includegraphics[width=6.5in]{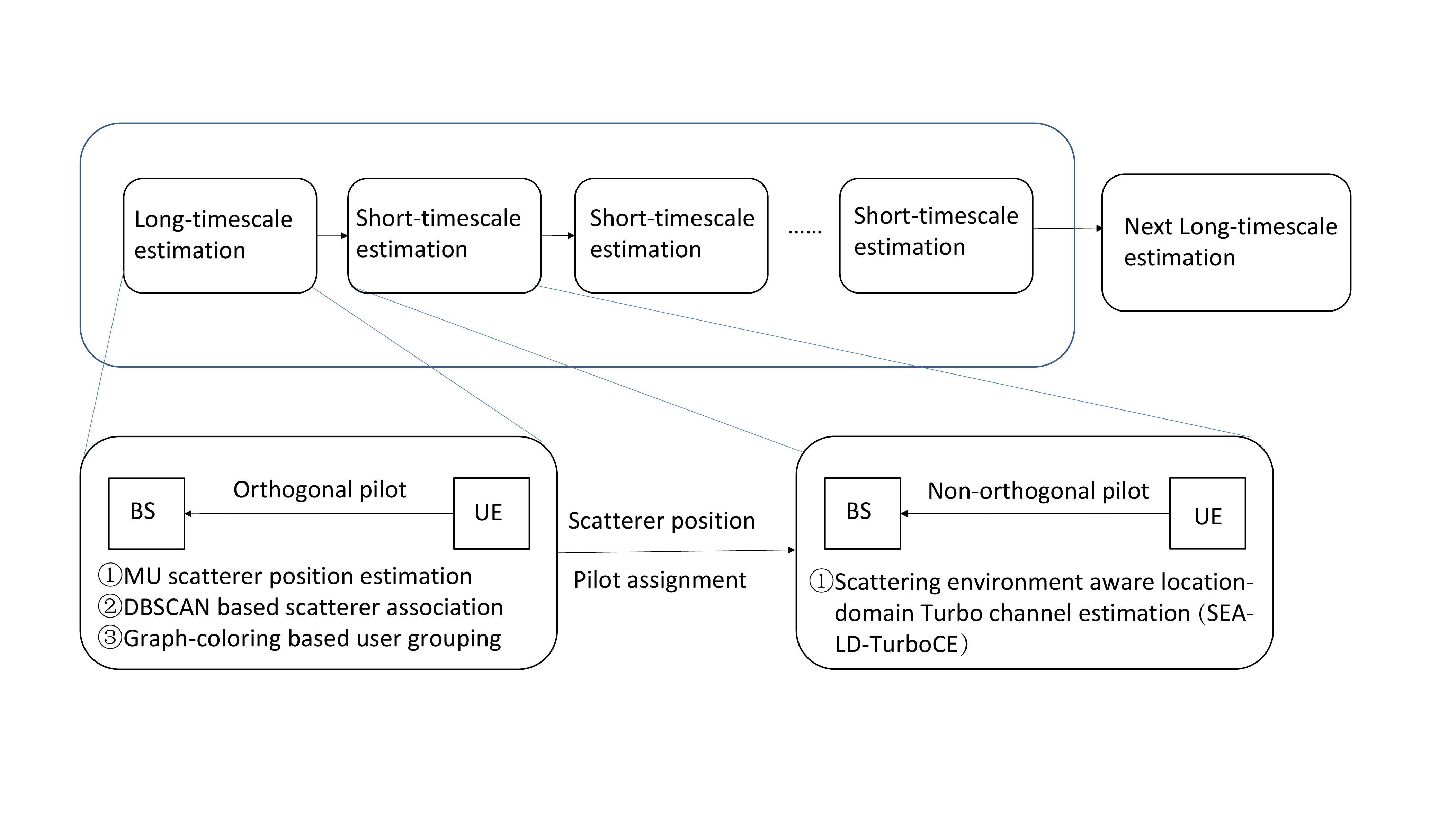}
      \caption{Two-timescale joint MU channel estimation and localization scheme.}
      \label{figure_1}
    \end{figure}

Specifically, for the long timescale phase, we employ the  TST-MUSIC algorithm to provide a coarse estimation of scatterer positions for each user. To enhance the accuracy of these positions, we associate the scatterers across users based on the DBSCAN algorithm. The BS then uses this information to apply a graph-coloring based user grouping algorithm, which enables spatial division multiplexing of pilots for users in the same group to reduce pilot overhead. Additionally, constructing a grid in the entire 3-D space  $\mathcal{R}$ leads to a typically large number of grid points \( Q \). With the algorithm's complexity being  \( \mathcal{O}(Q^3) \), the computational demands become prohibitively high. Thus, we leverage the prior information acquired by the long timescale to diminish \( Q \).  Specifically, for the short timescale phase, we initialize the dynamic grid $\boldsymbol{\gamma}$ by leveraging the prior information of scatterer positions acquired from the long timescale phase. The SEA-LD-TurboCE algorithm is introduced for joint channel estimation and localization under spatially reused pilots.

\subsection{Long-Timescale Problem Formulation}
\textbf{(1) Coarse scatterer localization:} Defining $\boldsymbol{{\theta}}_k=\{{\theta}_{l} | l \in \mathcal{L}_k \}$, $\boldsymbol{{\phi}}_k=\{{\phi}_{l} | l \in \mathcal{L}_k \}$, and $\boldsymbol{{\tau}}_{k}=\{{{\tau}_{{l}}\left( \boldsymbol{p}_{\mathrm{U\!E},k}\right)} | l \in \mathcal{L}_k \}$, we reformulate the received signal model (1) as:
\begin{equation}
\boldsymbol{Y}_k=\left[\boldsymbol{y}_{k, 1}, \ldots, \boldsymbol{y}_{k, P}\right]=\boldsymbol{A}\left(\boldsymbol{\theta}_k, \boldsymbol{\phi}_k\right) \boldsymbol{X}_k \boldsymbol{G}\left(\boldsymbol{\tau}_k\right)^T,
\end{equation}
where $\boldsymbol{A}\left(\boldsymbol{\theta}_k, \boldsymbol{\phi}_k\right)=\left[\boldsymbol{a}\left(\theta_{\mathrm{U\!E},k}, \phi_{\mathrm{U\!E},k}\right), \boldsymbol{a}\left(\theta_{1}, \phi_{1}\right), \ldots, \boldsymbol{a}\left(\theta_{L_k}, \phi_{L_k}\right)\right], \boldsymbol{X}_k=\operatorname{diag}\left(\boldsymbol{x}_k\right), \boldsymbol{G}(\boldsymbol{\tau}_k)=[\boldsymbol{g}\left({\tau}_{{0}}\left( \boldsymbol{p}_{\mathrm{U\!E},k}\right)\right), \boldsymbol{g}\left({\tau}_{{1}}\left( \boldsymbol{p}_{\mathrm{U\!E},k}\right)\right),\ldots, \boldsymbol{g}\left({\tau}_{{L_k}}\left( \boldsymbol{p}_{\mathrm{U\!E},k}\right)\right)]$ with $\boldsymbol{g}\left({\tau}_{{l}}\left( \boldsymbol{p}_{\mathrm{U\!E},k}\right)\right)=[u_{k, 1} e^{-j 2 \pi f_0 {\tau}_{{l}}\left( \boldsymbol{p}_{\mathrm{U\!E},k}\right)}, \ldots,$
$u_{k, P} e^{-j 2 \pi P f_0 {\tau}_{{l}}\left( \boldsymbol{p}_{\mathrm{U\!E},k}\right)}]^T.$
The scatterer localization problem of a single user $k$ can be formulated as: given the received signal $\boldsymbol{Y}_k$ and the user's location $\boldsymbol{p}_{\mathrm{U\!E},k}$, we need to estimate the scatterer's position ${\boldsymbol{\xi}}_k= \{{\boldsymbol{p}}_l | l \in \mathcal{L}_k\}$ in 3-D space.

\textbf{(2) Scatterer association and refined scatterer localization:}
Since MUs share the same scatterers, it is possible to associate the scatterers across users and then merge the locations of the shared scatterers to achieve a refined estimation of their locations. Specifically, to describe the association relationships between scatterers for MUs, we define a scatterer association vector \( \boldsymbol{\beta}=[\beta_1, \beta_2, \ldots, \beta_{\sum_{k=1}^{K} L_k}] \). Each element of $\boldsymbol{\beta}$ is an integer within the range of 0 to \( M \), which means this scatterer is assigned to $m$-th cluster, and $0$ means this scatterer does not belong to any cluster. Thus, the index set \( \mathcal{C}_m \) of cluster \( m \) is given by $ \mathcal{C}_m = \left\{ l| \beta_l = m, \forall l \right\}$. Besides, the error of scatterer position estimation obtained by TST-MUSIC for $k$-th user is assumed to follow a Gaussian distribution, specifically $(\boldsymbol{p}_l-\boldsymbol{\hat{p}}_l) \sim \mathcal{N}(0,\sigma_{l}^2\mathbf{I})\in \mathbb{C}^{3 \times 1}, l \in \mathcal{L}_k$, and the observations between different users are independent. Consequently, the problem is to obtain the scatterer association result \( \boldsymbol{\beta} \) and then refine the scatterer positions \( \hat{\boldsymbol{\Xi}}^{'} \) with given coarse estimation  \( \hat{\boldsymbol{\Xi}} = [\hat{\boldsymbol{\xi}}_1, \ldots,\hat{\boldsymbol{\xi}}_K] \). 

 \textbf{(3) User grouping for spatial pilot reuse:} If two users are located at a considerable distance from each other, the signal propagation paths between them may be significantly different. This difference leads to a higher level of orthogonality between their channels in the location domain. This orthogonality can be effectively utilized to enhance the capacity of pilots by allowing users in the same group to share the same pilots. The distance between two users is defined as the minimum distance between their respective primary scatterers, where primary scatterers refer to the paths that carry more than, say 95 percent of the total energy. A user grouping algorithm based on graph theory is used to ensure the scatterers corresponding to users do not overlap in spatial. The details are as follows.

 Assume there are $G$ orthogonal resource blocks, which are utilized by sharing in the time, frequency, and code domains. Thus, the $K$ users can be partitioned into $G$ groups, denoted as $\left\{\mathcal{G}_1, \ldots, \mathcal{G}_G\right\}$. Define the user grouping vector as $\boldsymbol{\omega}=[\omega_1, \ldots, \omega_K]$ and the elements in $\boldsymbol{\omega}$ is satisfying $\omega_{k} \in\{1,\ldots, G\}, \forall k\in  \mathcal{K}$, which indicates that the $k$-th user is allocated to the $g$-th group. Consequently, we have $\mathcal{G}_g = \{k \mid \omega_{k} = g, \forall k \in  \mathcal{K}\}$. The signals from different uses in one group are desired to be separated in the location domain and the minimum distance between all pairs of users in the same group is maximized, which can be represented as
\begin{equation}
\begin{aligned}
&\max _{\substack{\boldsymbol{\omega}}} \quad \min _{\substack{ k_1, k_2 \in \mathcal{G}_g},  k_1 \neq k_2} \quad d_{k_1, k_2}\\
\text { subject to }
&~\omega_{k_1}, \omega_{k_2} \in\{1,\ldots, G\}, \forall k_1, k_2 \in  \mathcal{K},
\end{aligned}
\end{equation}
where $d_{k_1, k_2}$ is the distance between two users, which is defined as the minimum distance between the positions of the scatterers associated with two users and can be obtained by
\begin{equation}
\hat{d}_{k_1, k_2}=\min _{l_{1} \in \mathcal{L}_{k_1}, l_{2} \in \mathcal{L}_{k_2} }|| \hat{\boldsymbol{p}}_{l_{1}}-\hat{\boldsymbol{p}}_{l_{2}}||,
\end{equation}
where $ \hat{\boldsymbol{p}}_{l_{1}}$ and $ \hat{\boldsymbol{p}}_{l_{2}}$ are the $l_{1}$-th and $l_{2}$-th  scatterer positions of user $k_1$ and $k_2$, respectively.
\subsection{Short-Timescale Problem Formulation}

In practice, the MU channels exhibit certain joint sparsity as explained in Section \ref{sec:system_model}. In this subsection, we shall introduce a three-layer sparse prior model to capture the joint sparse structure of the MU channels. Conditioned on the elements of channel support vectors $s_q^k$, the elements of $x_q^k$ are independent and the conditional prior distributions are given by
\begin{equation}
p\left(x_q^k \mid s_q^k\right)=\delta\left(s_q^k+1\right) \delta\left(x_q^k\right)+\delta\left(s_q^k-1\right) \mathcal{C N}\left(x_q^k ; 0,\left(\sigma_q^k\right)^2\right), q \in \mathcal{Q}.
\end{equation}
The joint distribution for the channel support vectors
\begin{equation}
p\left(\mathbf{s}, \mathbf{s}_1, \ldots, \mathbf{s}_K\right)=p(\mathbf{s}) \prod_{k \in \mathcal{K}} p\left(\mathbf{s}^k \mid \mathbf{s}\right)=p(\mathbf{s})\prod_{k \in \mathcal{K}} \prod_{q \in \mathcal{Q}} p\left(s_q^k \mid s_q\right),
\end{equation}
\begin{equation}
p\left(s_q^k \mid s_q\right)=\delta\left(s_q+1\right) \delta\left(s_q^k+1\right)+\delta\left(s_q-1\right)\left(\delta\left(s_q^k-1\right) \rho_{k}+\delta\left(s_q^k+1\right)(1-\rho_{k})\right),
\end{equation}
where $\rho_k = \frac{L_k}{L}$ represent the probabilities of $s^k_q = 1$ under the condition that $s_q = 1$, depicting the degree of overlap between individual scatterers and global scatterers. The joint support vector $s$ is modeled as a  i.i.d. prior:
\begin{equation}
p(\boldsymbol{s})= \prod_{q \in \mathcal{Q}} p\left( s_q\right),
\end{equation}
where the initialization distribution is set to $p\left(s_q=1\right)=\lambda$, and $\lambda$ determines the sparsity of the support vector. Thus, we have
\begin{equation}
p\left(\boldsymbol{s}, \boldsymbol{s}_1, \ldots,\boldsymbol{s}_K, \boldsymbol{x}_1, \ldots,\boldsymbol{x}_K\right)=p(\boldsymbol{s}) \prod_{k \in \mathcal{K}} \prod_{q \in \mathcal{Q}} p\left(s_q^k \mid s_q\right) \prod_{k \in \mathcal{K}} \prod_{q \in \mathcal{Q}} p\left(x_q^k \mid s_q^k\right).
\end{equation}

However, in the probability model (19) proposed above, the variables \( x_0^k \) are not included because their distribution models differ from the elements of \( \boldsymbol{x}_k \). For \( x_0^k \), we assume it follows a Bernoulli-Gaussian distribution with parameter \( \rho_0^k \), where \( \rho_0^k \) represents the probability of the presence of a LoS path in the communication link. A support vector \( s_0^k \in\{0,1\} \) is defined to indicate whether the communication link has a LoS path. If it exists, \( s_0^k = 1 \), otherwise, \( s_0^k = 0 \). The prior distribution \( p(s_0^k) \) follows a Bernoulli distribution, and the conditional distribution \( p(x_0^k \mid s_0^k) \) follows a Bernoulli-Gaussian distribution, as shown in the following equations:
\begin{equation}
p(s_0^k) = \delta(s_0^k - 1) \rho_0^k + \delta(s_0^k) (1 - \rho_0^k),
\end{equation}
\begin{equation}
p(x_0^k \mid s_0^k) = \delta(s_0^k) \delta(x_0^k) + \delta(s_0^k - 1) \mathcal{CN}(x_0^k; 0, (\sigma_0^k)^2).
\end{equation}

The joint prior distribution of all random variables is given by
\begin{equation}
\begin{aligned}
p &\left(\boldsymbol{s}, \boldsymbol{s}_1, \ldots, \boldsymbol{s}_K, \boldsymbol{x}_1, \ldots, \boldsymbol{x}_K, s_0^1, \ldots,  s_0^K, x_0^1, \ldots, x_0^K\right) \\
&=p\left(\boldsymbol{s}, \boldsymbol{s}_1, \ldots, \boldsymbol{s}_K, \boldsymbol{x}_1, \ldots, \boldsymbol{x}_K\right) \prod_{k \in \mathcal{K}} p\left(s_0^k\right)\prod_{k \in \mathcal{K}} p\left(x_0^k \mid s_0^k\right) .
\end{aligned}
\end{equation}


For convenience, we define two location domain sparse basis matrices in (20) and (21) for the NLOS and LOS paths, respectively. 
\begin{equation}
\boldsymbol{\Phi}_{Q, k}=\left[\begin{array}{c}
u_1 \boldsymbol{A}\left(\boldsymbol{\gamma}\right) \operatorname{diag}\left(\boldsymbol{b}_{k, 1}\left(\boldsymbol{\gamma}, \boldsymbol{p}_{U E, k}\right)\right) \\
\ldots \\
u_P \boldsymbol{A}\left(\boldsymbol{\gamma} \right) \operatorname{diag}\left(\boldsymbol{b}_{ k, P}\left(\boldsymbol{\gamma}, \boldsymbol{p}_{U E, k}\right)\right)
\end{array}\right],
\end{equation}
\begin{equation}
\boldsymbol{\Phi}_{0, k}=\left[\begin{array}{c}
u_1 e^{-j 2 \pi f_0\left({\tau}\left(\boldsymbol{p}_{U E, k}\right)\right)} \boldsymbol{a}\left({\theta}_{\mathrm{U\!E},k}, {\phi}_{\mathrm{U\!E},k}\right) \\
\ldots \\
u_P e^{-j 2 \pi P f_0\left({\tau}\left(\boldsymbol{p}_{U E, k}\right)\right)} \boldsymbol{a}\left({\theta}_{\mathrm{U\!E},k}, {\phi}_{\mathrm{U\!E},k}\right)
\end{array}\right].
\end{equation}
Due to the users within the same group transmit signals in the same time and frequency domains, the received signal of $g$-th group  $\boldsymbol{y_g}\in \mathbb{C}^{N P \times 1}$ can be expressed as
\begin{equation}
\boldsymbol{y_g}= \sum_{k \in \mathcal{G}_g}\left(\mathbf{\Phi}_k\left(\boldsymbol{\gamma}, \mathbf{p}_{\mathrm{U\!E}, k}\right) \boldsymbol{x}_{k}+\boldsymbol{z}_{{k}}\right),
\end{equation}
where $\boldsymbol{x}_k=\left[x^k_{0}, x^k_{1}, \ldots, x^k_{Q}\right]^T \in \mathbb{C}^{Q \times 1}$, $\boldsymbol{z}_k=\left[\left(\boldsymbol{z}_{k, 1}\right)^T, \ldots,\left(\boldsymbol{z}_{k, P}\right)^T\right]^T \in \mathbb{C}^{N P \times 1}$. $\boldsymbol{\Phi}_k\left(\boldsymbol{\gamma}, \boldsymbol{p}_{\mathrm{U\!E}, k}\right)=\left[\boldsymbol{\Phi}_{0, k}, \boldsymbol{\Phi}_{Q, k}\right] \in \mathbb{C}^{N P \times(Q+1)}$. Due to the propagation of users from different groups on orthogonal channels, the  received signal for all the users can be written as linear observation models with unknown parameters in measurement matrices:
\begin{equation}
\boldsymbol{y}=\left[\boldsymbol{y}_{\mathrm{1}}; \ldots ; \boldsymbol{y}_{G} \right]=\boldsymbol{\Phi} \boldsymbol{x}+\boldsymbol{z}\in \mathbb{C}^{GN P \times 1},
\end{equation}
where $\boldsymbol{\Phi}=\text { BlockDiag }\left(\boldsymbol{\bar {\Phi}}^{1}, \ldots, \boldsymbol{\bar {\Phi}}^{G}\right)\in \mathbb{C}^{G N P \times K(Q+1)}$, $\boldsymbol{\bar {\Phi}}^{g}=\left[\boldsymbol{\Phi}_k\left(\boldsymbol{\gamma}, \boldsymbol{p}_{U\!E, 1}\right), \ldots, \boldsymbol{\Phi}_k\left(\boldsymbol{\gamma}, \boldsymbol{p}_{U\!E, k}\right)\right]_{k\in \mathcal{G}_g} \in \mathbb{C}^{N P \times \left|\mathcal{G}_g\right|(Q+1)}$, $\boldsymbol{x}=\left[\boldsymbol{x}_{\mathrm{1}}; \ldots; \boldsymbol{x}_{K} \right]\in \mathbb{C}^{K(Q+1) \times 1}$, $\boldsymbol{y}=\left[\mathbf{y}_{\mathrm{1}}; \ldots; \mathbf{y}_{G} \right]\in \mathbb{C}^{GN P \times 1}$, and $\boldsymbol{z}=\left[\boldsymbol{z}_{\mathrm{1}}; \ldots; \boldsymbol{z}_{G} \right]\in \mathbb{C}^{GN P \times 1}$.

The task is to estimate the channel vector $\boldsymbol{x}$, the dynamic grid $\boldsymbol{\gamma}$, and the user position $\boldsymbol{P}_{\mathrm{U\!E}}=[\boldsymbol{p}_{\mathrm{U\!E},1}, \ldots, \boldsymbol{p}_{\mathrm{U\!E},k}]$ through the given $\boldsymbol{y}$ and the prior information of scatterer positions  $\hat{\boldsymbol{\Xi}}^{'}$. On one hand, with the given dynamic grid $\boldsymbol{\gamma}$ and user position $\boldsymbol{P}_{\mathrm{U\!E}}$, our goal is to calculate the conditional marginal posterior probabilities $ p\left(x_{{q}}^k \mid \boldsymbol{y}, \boldsymbol{\zeta}, \hat{\boldsymbol{\Xi}^{'}}\right), \quad p\left(s_q^k \mid \boldsymbol{y}, \boldsymbol{\zeta}, \hat{\boldsymbol{\Xi}^{'}}\right),$ for $\forall q \in \mathcal{Q}_k, \forall k \in \mathcal{K}$ with $\boldsymbol{\zeta} = \{\boldsymbol{\gamma}, \boldsymbol{P}_{\mathrm{U\!E}}\}$ is the uncertainty parameters. On the other hand, the dynamic grid $\boldsymbol{\gamma}$ and user positions $\boldsymbol{P}_{\mathrm{U\!E}}$ can be obtained through MAP estimation:
\begin{align}
\boldsymbol{\hat{\zeta}} = \arg \max_{\boldsymbol{\zeta}} \ln p\left(\boldsymbol{\zeta} \mid \boldsymbol{y}\right), 
\end{align}
However, the presence of loop structures in the underlying probability model's factor graph makes the exact computation of the aforementioned conditional marginal posterior probabilities very challenging. In the next section, we propose a SEA-LD-TurboCE algorithm, which approximates the computation of conditional marginal posterior probabilities through a turbo framework combining linear minimum mean square error (LMMSE) estimation and message passing. It also employs the expectation-maximization (EM) method to find an approximate solution to (24).
\section{Long-Timescale Algorithm}\label{sec:Long-Timescale Algorithm}

\subsection{TST-MUSIC Algorithm for Coarse Scatterer Localization}
We decompose the problem into the angle/delay estimation and the range estimation using the delay of scatterers and the user's position. For the $k$-th user, $\boldsymbol{{\theta}_k}, \boldsymbol{{\phi}}_k$, and $\boldsymbol{{\tau}}_{k}$ can be estimated by the existing algorithm, such as TST-MUSIC \cite{ref26}. This algorithm combines the temporal filtering techniques and the spatial beamforming techniques to jointly estimate the angles and the delays of the multipath, and the details are omitted here. Then we need to ulize $[\boldsymbol{\hat{\theta}}_k,\boldsymbol{\hat{\phi}}_k, \boldsymbol{\hat{\tau}}_k]$ and $\boldsymbol{p}_{\mathrm{U\!E},k}$ of the $k$-th user to estimate the scatterer positions $\hat{\boldsymbol{\xi}}_k=\{[\hat{\theta}_{l}, \hat{\phi}_{l}, {\hat{r}_{{l}}}] | l \in \mathcal{L}_k\}$ with $\hat{r}_{l}$ denotes the estimation of the range of the scatterer \(l\). Thus, the scatterer range estimation problem can be reformulated by minimizing the following function:
\begin{equation}
\min_{r_{l}} \|\frac{\sqrt{{r_{\mathrm{U\!E},k}^2+r_{l}^2-2 r_{\mathrm{U\!E},k} r_{l}}\hat{\alpha}_k}}{c} +\frac{r_{l}}{c}-{{\tau}_{{l}}\left( \boldsymbol{p}_{\mathrm{U\!E},k}\right)}\|^2, l \in \mathcal{L}_k,
\end{equation}
where $\hat{\alpha}_k=\left[\sin \left(\theta_{\mathrm{U\!E},k}\right) \sin \left(\hat{\theta}_{l}\right) \cos \left(\phi_{\mathrm{U\!E},k}-\hat{\phi}_{l}\right)+\cos \left(\theta_{\mathrm{U\!E},k}\right) \cos \left(\hat{\theta}_{l}\right)\right]$. This problem can be solved by existing methods, such as the Newton-Raphson method \cite{ref29}, and is omitted here for simplicity.

\subsection{DBSCAN Based Scatterer Association for Refined Scatterer Localization}
As described in the MU joint sparsity model, the set of scatterers for each user is a subset of the global scatterers set.  Consequently, it is feasible to reconstruct the scattering environment by utilizing the scatterer subsets observed by MU.  This requires identifying and associating shared scatterers of MUs.  Thus, we propose a DBSCAN-based algorithm aimed at achieving scattering environment awareness as well as joint channel estimation and localization for MU. 

DBSCAN is renowned for its effectiveness in noise elimination and its adept handling of well-separated observations, making it an ideal choice for scatterer association\cite{ref31}. It hinges on two crucial initial parameters: the neighborhood radius \( \varepsilon \) and the minimum number of points $MinPts$. $\varepsilon$ defines the radius of the neighborhood around a point, used to determine the density of scatterers, where neighborhood refers to the Euclidean distance between two scatterers. $MinPts$ is the minimum number of scatterers that must be present within a region for it to be classified as a cluster.  A point selected at random is deemed a core point if it can claim at least $MinPts$ points within \( \varepsilon \). Those points that fall within the neighborhood radius of the core point and meet this criterion are collectively labeled, thus constituting a cluster. 

The DBSCAN-based scatterer association process can be described as follows: 1) Select a scatterer $\hat{\boldsymbol{\Xi}}[i]$ as a sample with $\hat{\boldsymbol{\Xi}}[i]$ is the $i$-th row of  $\hat{\boldsymbol{\Xi}}$. If the neighborhood within $\varepsilon$ contains more than $ MinPts $ points, a new cluster $ \mathcal{C} $ is initialized, and the scatterer sample $ \hat{\boldsymbol{\Xi}}[i] $ is marked as visited.  2) Iterate through all the scatterers in the cluster $ \mathcal{C} $. If any scatterer's $ \varepsilon $-neighborhood contains more than $ MinPts $ scatterers, add all the scatterers to cluster $ \mathcal{C} $ and mark them as visited. 3) Repeat steps 1 and 2 until all the density-connected scatterers are identified for cluster $ \mathcal{C} $. 4) Check for a new unvisited scatterer and repeat steps 1-3 to identify additional clusters. 5) Finally, label all the unvisited scatterers as noise samples and output \( \boldsymbol{\beta} \). Note that the individual scatterers are treated as noise because they are unique to a single observation and do not belong to any cluster \( \mathcal{C}_m \). Besides, the shared scatterers are common to multiple observations and thus can be effectively identified. According to the assumption of the error of coarse scatterer localization, the position of the shared scatterer in cluster \( m \) is refined by 
\begin{equation}
 \bar{\boldsymbol{\xi}}_m = \frac{1}{|\mathcal{C}_m|} \sum_{i \in \mathcal{C}_m}\hat{\boldsymbol{\Xi}}[i],
\end{equation}
where \( |\mathcal{C}_m| \) represents the number of elements in the set \( \mathcal{C}_m \). Thus, the all shared scatterers is given by $\boldsymbol{\Xi}_{\text{shared}} = \begin{bmatrix}
\bar{\boldsymbol{\xi}}_{1}, \ldots, \bar{\boldsymbol{\xi}}_{M}
\end{bmatrix}\in \mathbb{C}^{M \times 3}$, and the refined  positions of shared scatterer are represented as:
\begin{equation}
\hat{\boldsymbol{\Xi}}^{'}[i]=
\begin{cases} 
 \bar{\boldsymbol{\xi}}_m & \text{if } i \in \mathcal{C}_m, \\
\hat{\boldsymbol{\Xi}}[i] & \text{otherwise}.
\end{cases}
\end{equation}
Following the aforementioned process, we can obtain the updated global scatterer positions as well as the associations between scatterers for MUs.  To avert the severe performance degradation caused by missed detections, the grid $\boldsymbol{\gamma}_k$ should encompass all possible scatterer positions for the $k$-th user.  Therefore, the dynamic grid is initialized as $\boldsymbol{\gamma}^0= \hat{\boldsymbol{\Xi}}^{'}$.

\subsection{Graph-Coloring Based User Grouping}
The optimization problem presented in (10) is an NP-hard mixed-integer programming problem, which implies that there is no polynomial-time algorithm to find an optimal solution. Therefore, we transform the user grouping problem into a graph coloring problem and use the PSO algorithm to find a suboptimal solution. Specifically, an undirected unweighted graph is defined by an adjacency matrix \( \boldsymbol{E} \in \mathbb{C}^{K \times K} \), and can be represented as:
 \begin{equation}
\boldsymbol{E} = \begin{bmatrix}
0      & E_{12} & \ldots & E_{1K} \\
E_{21} & 0      & \ldots & E_{2K} \\
\vdots & \vdots & \ddots & \vdots \\
E_{K1} & E_{K2} & \ldots & 0
\end{bmatrix}.
 \end{equation}
where each element \( E_{ij}, i \neq j \) is defined as:
\begin{equation}
E_{ij} =
\begin{cases}
1, & \text{if } d_{i,j} < d_{\text{adj}}, \\
0, & \text{if } d_{i,j}  \geq d_{\text{adj}},
\end{cases}
\end{equation}
where \( d_{i,j}  \) denotes the distance between vertex \( i \) and vertex \( j \) defined by (11), and $d_{\text{adj}}$ is the adjacency distance and can be adjusted to ensure the users within the same group are quasi-orthogonal within the location domain. This equation means that if the distance between vertex \( i \) and vertex \( j \) is less than \( d_{\text{adj}} \), then there is an edge between these two users. Besides, the diagonal elements \( E_{ii} \) are always \( 0 \), as there is no edge connecting a vertex to itself. 

Next, we introduce how to obtain the user grouping vector $\boldsymbol{\omega}$ based on the PSO method \cite{ref28}.  This heuristic methodology conceptualizes each entity as a particle navigating through a multi-dimensional search expanse at a defined velocity. Specifically, the $k$-th element of $\boldsymbol{\omega}$ is defined as the position of the particle in the $t$-th iteration, denoted as $X_{a,k}^{(t)}$,  where $k \in \mathcal{K}$, $a \in \{1,2, \ldots, A_p\}$ with $A_p$ denoting the number of particles. Similarly, $V_{a,k}^{(t)}$ is defined to represent the velocity of the particle. The particle's velocity and position are iterated according to the evolutionary rules given by the particle swarm algorithm, thus forming a coloring scheme for the entire graph $G$. Specifically, in the $t$-th iteration, the velocity and position of each particle are updated within the integer search space based on the following two equations:
\begin{equation}
V_{a,k}^{(t+1)}=\text{int}(V_{a,k}^{(t)})+Z_1+Z_2,  
\end{equation}
\begin{equation}
X_{a,k}^{(t+1)}=X_{a,k}^{(t)}+V_{a,k}^{(t+1)},
\end{equation}
where $X_{a,k}^{(t)}$ and $V_{a,k}^{(t)}$ are the position and velocity of the $k$-th element of the $a$-th particle, respectively, and the notation \(\text{int}(\cdot)\) signifies the rounding operation. Besides, $Z_1$ and $Z_2$ are the random integer uniformly distributed within the interval $[0, c_1({pbest}_{a,k} - X_{a,k}(t))]$ and $[0, c_2({gbest}_{a,k} - X_{a,k}(t))]$, respectively, with $\boldsymbol{gbest}^{(t)}=\{gbest_{k}^{(t)}\}_{k=1}^K$ denotes the best position in the whole swarm, $pbest_{a,k}^{(t)}$ denotes the best position values of $X_{a,k}^{(t)}$, and $c_1$ and $c_2$ are acceleration coefficients. To meet the following two constraints: 1) Adjacent vertices in the graph must be colored differently. 2) The number of colors used should be as few as possible, the fitness function 
can be defined as:
\begin{equation}
f_\text{fitness} = \sum_{i=1}^K \sum_{j=1}^K E_{ij} S_{ij} + \max(g-G,0),
\end{equation}
where \( S_{ij} \) is used to indicate whether vertex \( i \) and vertex \( j \) are colored the same, i.e., if  $ X_{a,i} =  X_{a,j}$ then $S_{ij} =  1$, otherwise $S_{ij} =  0$, and \(g\) represents the actual number of colors used for coloring. For each iteration, 
$pbest_{a, k}^{(t+1)}$ and $gbest_k^{(t+1)}, \forall a, k.$ need to update if $f_\text{fitness}$ become smaller, and the user grouping algorithm operates iteratively with a computational complexity of \( \mathcal{O}(K^2) \) per iteration until the number of color \( g \) equal to the count of user groups \( G \). The algorithm is summarized in Algorithm 1. To select an appropriate initial distance \( d_{adj}^{(0)} \), one must balance the intra-group interference and convergence speed. A logical approach is to set \( d_{adj}^{(0)} \) as the \( e \)-th percentile of the cumulative distribution function (CDF) of user distances. \( e \) should adjust with the number of groups \( G \): a smaller \( e \) for smaller \( G \) (more users per group) and a larger \( e \) for larger \( G \) (fewer users per group). Therefore, we set \( e = \frac{G}{K} \), allowing \( d_{adj}^{(0)} \) to adapt to different group sizes. Finally, utilizing the optimal pilot assignment $\boldsymbol{\omega}$, we allocate the same pilots to all users within the same group during subsequent short-timescale phases, resulting in the use of non-orthogonal pilots.
\begin{algorithm}[H]
\caption{PSO-based Graph Coloring Algorithm for User Grouping.}
\label{alg:graph_coloring}
\begin{algorithmic}[1]
\STATE \textbf{Input:} the scatterers for all users $\hat{\boldsymbol{\xi}}_k, \forall k \in \mathcal{K}$, the number of groups $G$, the update step of adjacency distance $\Delta d_{adj}$, and the threshold $\epsilon$.
\STATE  Initialize the adjacency matrix $\boldsymbol{E}$ by (28) with $d_{adj}^{(0)}$, $X_{a,k}^{(0)} = k$, $V_{a,k}^{(0)}, gbest_k^{(0)}$, ${pbest}_{a,k}^{(0)}, \forall a,k$, and 
$t=0$.
\WHILE{$g>G$}
\WHILE{$\|\boldsymbol{gbest}^{(t+1)}-\boldsymbol{gbest}^{(t)}\|\geq \epsilon$} 
       \STATE     Update the velocity of particle $\forall k \in\mathcal{K}$ using (30).
        \STATE    Update the position of particle $\forall k \in \mathcal{K}$ using (31).
       \STATE     Calculate the fitness of particle $\forall k \in \mathcal{K}$ based on the objective in (32).
       \STATE Update $pbest_{a, k}^{(t+1)}$ and $gbest_k^{(t+1)}, \forall a, k.$ if $f_\text{fitness}^{(t+1)}<f_\text{fitness}^{(t)}$.
       \STATE  $t=t+1$.
\ENDWHILE
\STATE  Update the the adjacency matrix $\boldsymbol{E}$ with $d_{adj}^{(t+1)} = d_{adj}^{(t)} - \Delta d_{adj}$ by (28).
\ENDWHILE
\STATE \textbf{Output:} Pilots assignment $\boldsymbol{\omega} = \boldsymbol{gbest}$.
\end{algorithmic}
\end{algorithm}


\section{Short-Timescale Algorithm}\label{sec:DBSCAN-Turbo-SBI Algorithm}

Based on the prior information of MUs obtained by the refined scatterer positions, our objective is to attain an accurate joint channel estimation and localization of MUs based on the received pilots at the current time and the prior information in the long timescale estimation. Inspired by the EM method \cite{ref27}, the SEA-LD-TurboCE algorithm performs iterations between the following two steps until convergence.

\textbf{E Step:} For given uncertainty parameters $\boldsymbol{\zeta}^t$ in the $t$-th iteration, calculate the approximate posterior $p\left(\boldsymbol{x} \mid \boldsymbol{y}, \boldsymbol{\zeta}^t\right)$ by message passing method via turbo framework.

\textbf{M Step:} Using the approximate posterior $p\left(\boldsymbol{x}\mid \boldsymbol{y}, \boldsymbol{\zeta}^t\right)$ obtained in the SEA-LD-TurboCE-E step,
 construct a surrogate function for function $\ln p(\boldsymbol{y} \mid \boldsymbol{\zeta})$, then maximize the surrogate function with respect to $\boldsymbol{\zeta}$  via gradient ascent method to update the next iterate $\boldsymbol{\zeta}^{t+1}$.

The E Step contains two modules, as shown in Fig.3. Module A is an LMMSE estimator based on the observation $\boldsymbol{y}$ and extrinsic messages from Module B. Module B, performs MMSE estimation that combines the i.i.d. prior and the extrinsic messages from Module A. The two modules are executed iteratively until convergence.
   \begin{figure}[t]
      \centering
      \includegraphics[width=4.5in]{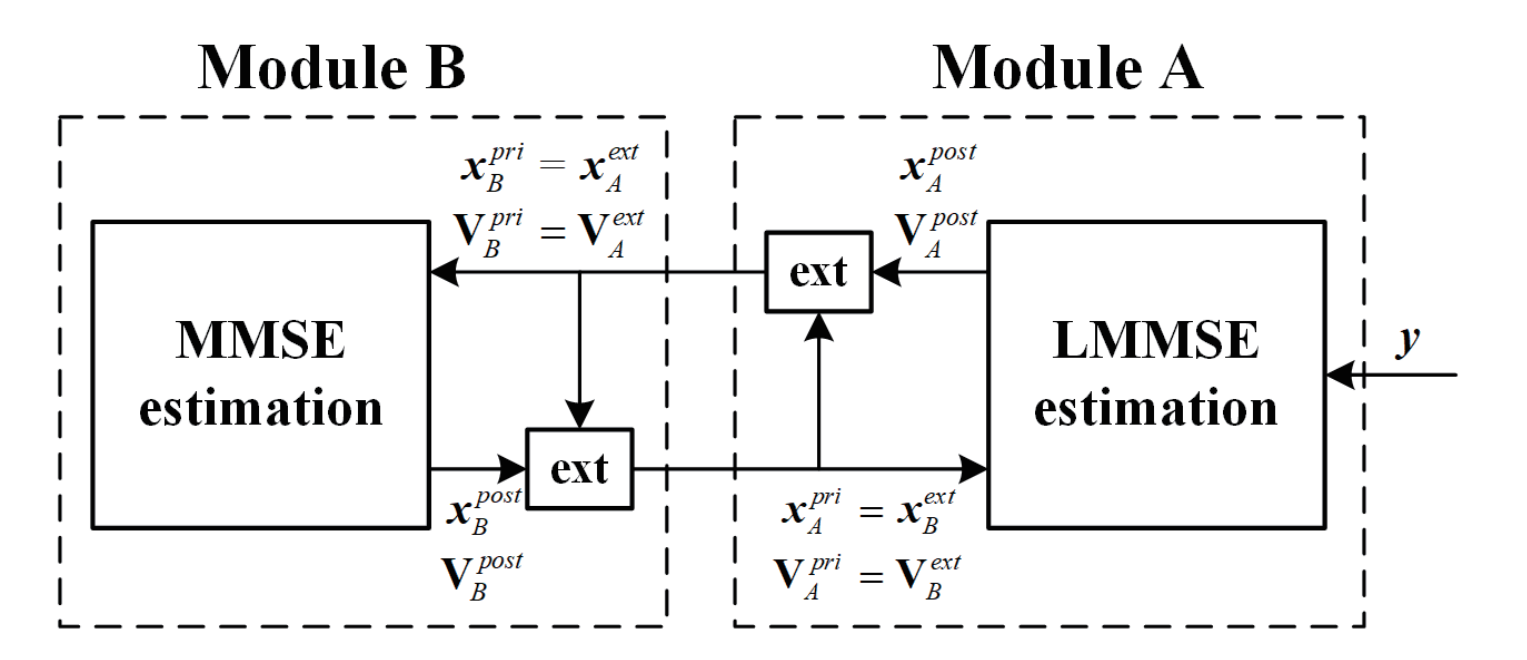}
      \caption{ Modules of the SEA-LD-TurboCE E step and message flow between different modules.}
      \label{figure_1}
    \end{figure}
A. LMMSE in Module A:1) We assume that the prior distribution of $\boldsymbol{x}$ is $\mathcal{C N}\left(\boldsymbol{x} ; \boldsymbol{x}_A^{p r i}, \mathbf{V}_A^{p r i}\right)$, where $\boldsymbol{x}_A^{p r i}$ and $\mathbf{V}_A^{p r i}$ are the extrinsic mean and covariance matrix from Module $\mathrm{B}$, respectively. The posterior mean and covariance matrix of the LMMSE estimation are respectively given by
\begin{equation}
 \begin{aligned}
\boldsymbol{x}_A^{\text {post }} & =\mathbf{V}_A^{\text {post }}\left(\left(\mathbf{V}_A^{\text {pri }}\right)^{-1} \boldsymbol{x}_A^{\text {pri }}+\frac{\boldsymbol{\Phi}^H \boldsymbol{y}}{\sigma_z^2}\right), \\
\mathbf{V}_A^{\text {post }} & =\left(\frac{\boldsymbol{\Phi}^H \boldsymbol{\Phi}}{\sigma_z^2}+\left(\mathbf{V}_A^{\text {pri }}\right)^{-1}\right)^{-1} .
\end{aligned}   
\end{equation}
By subtracting the prior information from the posterior information, we obtain the extrinsic message from Module A as follows:
\begin{equation}
\begin{aligned}
\boldsymbol{x}_{A \rightarrow B}^{\text {ext }} & =\mathbf{V}_A^{\text {ext }}\left(\left({\mathbf{V}}_A^{\text {post }}\right)^{-1} \boldsymbol{x}_A^{\text {post }}-\left(\mathbf{V}_A^{\text {pri }}\right)^{-1} \boldsymbol{x}_A^{\text {pri }}\right), \\
\mathbf{V}_{A \rightarrow B}^{\text {ext }} & =\left(\left({\mathbf{V}}_A^{\text {post }}\right)^{-1}-\left(\mathbf{V}_A^{\text {pri }}\right)^{-1}\right)^{-1},
\end{aligned}
\end{equation}
   \begin{figure}[t]
      \centering
      \includegraphics[width=4.5in]{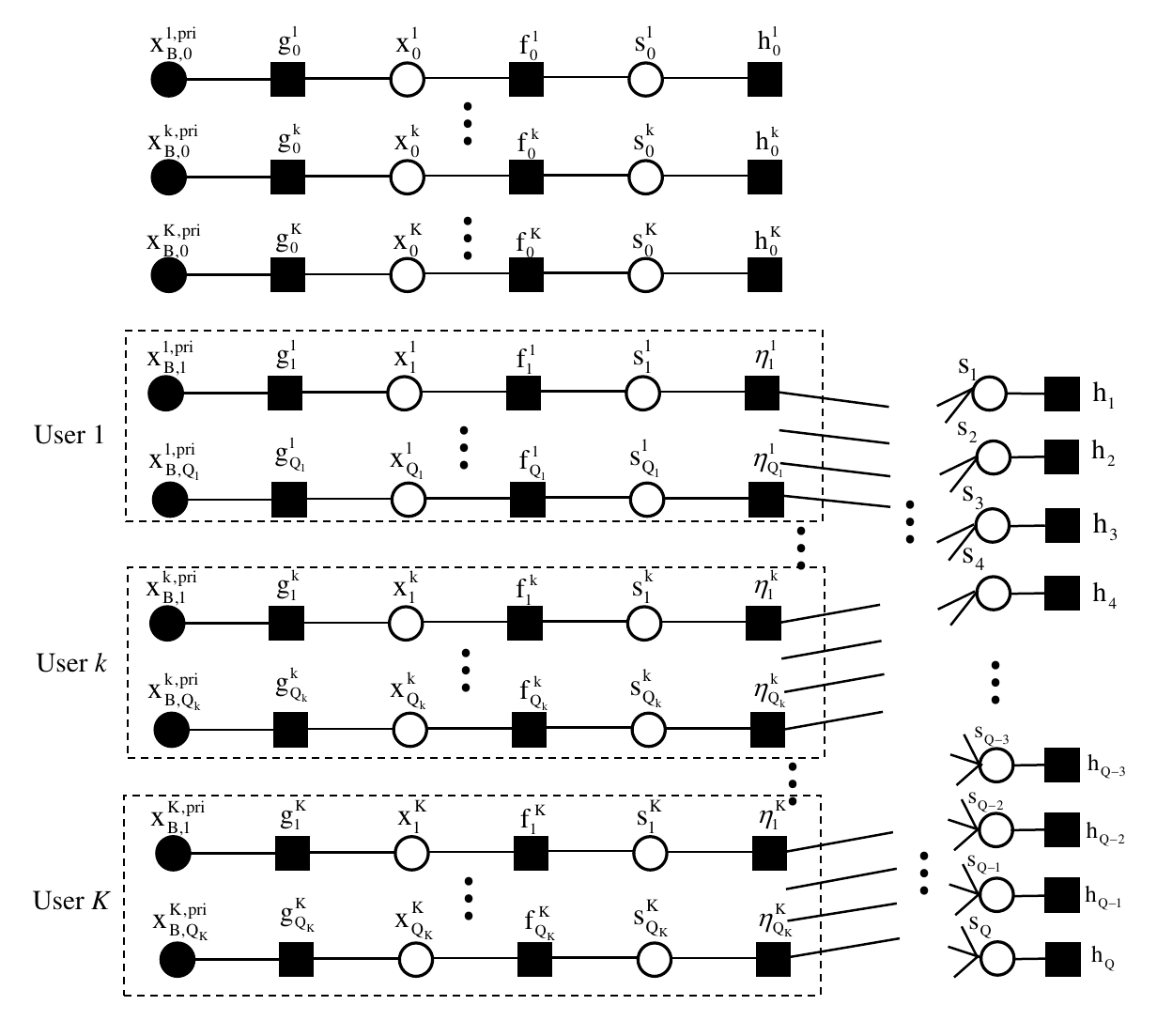}
      \caption{Factor graph of the joint distribution.}
      \label{figure_1}
    \end{figure}
B. Message passing in Module B: In Module B, a message passing scheme is used for the HMM-MMSE estimator to calculate the posterior of $\boldsymbol{x}$ and $\boldsymbol{s}$, based on the i.i.d channel prior and the extrinsic messages $\boldsymbol{x}_{A \rightarrow B}^{e x t}, \boldsymbol{V}_{A \rightarrow B}^{e x t}$ from Module A.  Specifically, the extrinsic messages $\boldsymbol{x}_{A \rightarrow B}^{e x t}, \boldsymbol{V}_{A \rightarrow B}^{e x t}$ are equivalently modeled as a virtual AWGN observation model:
\begin{equation}
\boldsymbol{x}_B^{pri}=\boldsymbol{x}+\mathbf{z}, 
\end{equation}
where $\boldsymbol{x}_B^{pri}=\boldsymbol{x}_{A \rightarrow B}^{ext}=\left[\left(\boldsymbol{x}_B^{1, p r i}\right)^T, \ldots, \left(\boldsymbol{x}_B^{K, pri}\right)^T\right]^T$ is treated as observations obtained via a virtual AWGN channel with zero mean noise vectors $\mathbf{z}^1, \ldots, \mathbf{z}^K$ and the extrinsic covariance $\boldsymbol{V}_{A \rightarrow B}^{e x t}=\operatorname{BlockDiag}\left(\boldsymbol{V}_B^{1, \text {pri }}, \ldots, \boldsymbol{V}_B^{K, \text {pri }}\right)$ is treated as the noise covariance, i.e., $\mathbf{z}^k \sim \mathcal{C N}\left(0; \boldsymbol{V}_B^{k, p r i}\right)$. The factor graph $\mathcal{G}_B$ of the joint distribution associated with this virtual AWGN observation model is shown in Fig. 4, where the function expression of each factor node is listed in Table I. In Table I, $x_{B, m}^{k, p r i}$ is the $m$-th elements of $\boldsymbol{x}_B^{k, p r i}$, and $v_{B, m}^{k, p r i}$ is the $m$-th diagonal elements of $\boldsymbol{V}_B^{k, p r i}$.
$$
\begin{array}{|c|c|c|}
\hline \text { Factor } & \text { Distribution } & \text { Functional form } \\
\hline \hline 
g_0^k\left(x_{B, 0}^{k, p r i}, x_0^k\right) & p\left(x_0^k \mid x_{B, 0}^{k, p r i}\right) & \mathcal{C N}\left(x_0^k ; x_{B, 0}^{k, p r i}, v_{B, 0}^{k, p r i}\right) \\

g_q^k\left(x_{B, q}^{k, p r i}, x_q^k\right) & p\left(x_q^k \mid x_{B, q}^{k, p r i}\right) & \mathcal{C N}\left(x_q^k ; x_{B, q}^{k, p r i}, v_{B, q}^{k, p r i}\right) \\
\hline \hline 
f_0^k\left(x_0^k, s_0^k\right) & p\left(x_0^k \mid s_0^k\right) & \mathcal{C N}\left(x_q^k ; x_{B, q}^{k, p r i}, v_{B, q}^{k, p r i}\right) \\
 f_q^k\left(x_q^k, s_q^k\right) & p\left(x_q^k \mid s_q^k\right) & \delta\left(s_q^k\right) \delta\left(x_q^k\right)+\delta\left(s_q^k-1\right) \mathcal{N}\left(x_q^k ; 0,\left(\sigma_q^k\right)^2\right) \\

\hline \hline \eta_q^k\left(s_q^k, s_q\right) & p\left(s_q^k \mid s_q\right) & \delta\left(s_q\right) \delta\left(s_q^k\right)+\delta\left(s_q-1\right)\left(\delta\left(s_q^k-1\right) \rho_k+\delta\left(s_q^k\right)\left(1-\rho_k\right)\right) \\

\hline \hline h_0^k\left(s_0^k\right) & p\left(s_0^k\right) & \delta\left(s_0^k-1\right) \rho_0^k+\delta\left(s_0^k\right)\left(1-\rho_0^k\right) \\

 h_q\left(s_q\right) & p\left(s_q\right) & \delta\left(s_q-1\right) \lambda+\delta\left(s_q\right)(1-\lambda) \\
\hline
\end{array}
$$

We now outline the message passing scheme over the factor graph $ \mathcal{G}_B$. The details are elaborated in Appendix A. According to the sum-product rule, we calculate the message passing on the factor graph depicted. Initially, we infer the messages traversing the paths \( x_q^k \rightarrow f_q^k \rightarrow s_q^k \rightarrow \eta_q^k \rightarrow s_q \), followed by the independent computation of messages along the paths \( x_0^k \rightarrow f_0^k \rightarrow s_0^k \rightarrow h_0^k \). Subsequently, the bidirectional message passing is executed by calculating \( s_q \rightarrow h_q \) and \( h_q \rightarrow s_q \). Finally, messages are fed back through the paths \( s_q \rightarrow \eta_q^k \rightarrow s_q^k \rightarrow f_q^k \rightarrow x_q^k \), \( h_0^k \rightarrow s_0^k \rightarrow f_0^k \rightarrow x_0^k \).

After calculating the updated messages $\left\{v_{\boldsymbol{f}_q^k \rightarrow \boldsymbol{x}_q^k}\right\}$, the approximate posterior distributions are given by
\begin{equation}
\hat{p}\left(x_q^k \mid \boldsymbol{y}\right) \propto v_{\boldsymbol{f}_q^k \rightarrow \boldsymbol{x}_q^k} \times v_{\boldsymbol{x}_q^k \rightarrow \boldsymbol{f}_q^k}, 
\end{equation}
\begin{equation}
\hat{p}\left(s_q^k \mid \boldsymbol{y}\right)=\frac{\pi_{s^k, q}^{i n} \pi_{s^k, q}^{\text {out }}}{\pi_{s^k, q}^{i n} \pi_{s^k, q}^{\text {out }}+\left(1-\pi_{s^k, q}^{\text {in }}\right)\left(1-\pi_{s^k, q}^{\text {out }}\right)}, \forall q \in \mathcal{Q}_k, \forall k \in \mathcal{K},
\end{equation}
where $v_{\boldsymbol{f}_q^k \rightarrow \boldsymbol{x}_q^k}, v_{\boldsymbol{x}_q^k \rightarrow \boldsymbol{f}_q^k}, \pi_{\boldsymbol{s}^k, q}^{i n}, \pi_{\boldsymbol{s}^k, q}^{o u t}$ are given in Appendix A. Then the posterior mean $\boldsymbol{x}_B^{k, p o s t}=$ $\left[x_{B, 1}^{k, \text { post }}, \ldots, x_{B, Q}^{k, \text { post }}\right]^T$ and variance $\boldsymbol{V}_B^{k, \text { post }}=\operatorname{Diag}\left(\left[v_{B, 1}^{\text {post }}, \ldots, v_{B, Q}^{\text {post }}\right]\right)$ for $\boldsymbol{x}^k$ can be calculated, respectively, as

\begin{equation}
x_{B, q}^{k, \text { post }}=\int_{x_q^k} x_q^k \hat{p}\left(x_q^k \mid \boldsymbol{y}\right), \\
v_{B, q}^{\text {post }}=\int_{x_q^k}\left|x_q-x_{B, q}^{k, p o s t}\right|^2 \hat{p}\left(x_q^k \mid \boldsymbol{y}\right), \forall q \in \mathcal{Q}_k, \forall k \in \mathcal{K}.
\end{equation}

Then the extrinsic message passed to Module A can be calculated as
\begin{equation}
\begin{aligned}
\boldsymbol{V}_{B \rightarrow A}^{\text {ext }}  &
=\left(\left(\boldsymbol{V}_B^{\text {post }}\right)^{-1}-\left(\boldsymbol{V}_B^{\text {pri }}\right)^{-1}\right)^{-1}, \\
\boldsymbol{x}_{B \rightarrow A}^{\text {ext }}  &=\boldsymbol{V}_{B \rightarrow A}^{\text {ext }}\left(\left(\boldsymbol{V}_B^{\text {post }}\right)^{-1} \boldsymbol{x}_B^{\text {post }}-\left(\boldsymbol{V}_B^{\text {pri }}\right)^{-1} \boldsymbol{x}_B^{\text {pri }}\right),
\end{aligned}
\end{equation}
where $\boldsymbol{x}_B^{\text {pri }}=\left[\left(\boldsymbol{x}_B^{1, \text {pri}}\right)^T, \ldots, \left(\boldsymbol{x}_B^{K, \text {pri}}\right)^T\right]^T, \boldsymbol{x}_B^{\text {post }}=\left[\left(\boldsymbol{x}_B^{1,\text {post }}\right)^T, \ldots, \left(\boldsymbol{x}_B^{K, \text {post }}\right)^T\right]^T, \boldsymbol{V}_B^{\text {pri}} \triangleq \operatorname{BlockDiag}\left(\boldsymbol{V}_B^{1, \text {pri }}, \ldots, \boldsymbol{V}_B^{K, \text {pri }}\right)$ and $\boldsymbol{V}_B^{\text {post } \triangleq} \triangleq \operatorname{BlockDiag}\left(\boldsymbol{V}_B^{1, \text {post }}, \ldots, \boldsymbol{V}_B^{K, \text {post }}\right)$.

In M Step, due to the difficulty in finding a closed-form expression for $\ln p(\boldsymbol{y}, \boldsymbol{\zeta})$, it is challenging to directly solve for the maximization of the log-posterior function in (30) is challenging. To address this issue, a common approach is to construct a surrogate function of $\ln p(\boldsymbol{y}, \boldsymbol{\zeta})$ and maximize this surrogate function instead with respect to $\boldsymbol{\zeta}$. Specifically, in the $t$-th iteration, the surrogate function constructed by the EM method is given by

\begin{equation}
\begin{aligned}
Q\left(\boldsymbol{\zeta} ; \boldsymbol{\zeta}^t\right) & =\int_{\boldsymbol{x}} p\left(\boldsymbol{x} \mid \boldsymbol{y}, \boldsymbol{\zeta}^t\right) \ln \frac{p(\boldsymbol{y}, \boldsymbol{x}\mid \boldsymbol{\zeta})}{p\left(\boldsymbol{x}\mid \boldsymbol{y}, \boldsymbol{\zeta}^t\right)} \\
& =-\left\|\boldsymbol{y}-\boldsymbol{\Phi} \boldsymbol{x}^{\text {post }}\right\|^2+\operatorname{tr}\left(\boldsymbol{\Phi} \mathbf{V}^{\text {post }} \boldsymbol{\Phi}^H\right)+C,
\end{aligned}
\end{equation}
where $C$ is a constant, $\boldsymbol{x}^{\text {post }}$ and $\mathbf{V}^{\text {post }}$ is the posterior mean and covariance of $\boldsymbol{x}$ obtained in the E step. In the current iteration $\boldsymbol{\zeta}^t$, the surrogate function $Q\left(\boldsymbol{\zeta} ; \boldsymbol{\zeta}^t\right)$ and its gradient satisfy the following properties: $Q\left(\boldsymbol{\zeta} ; \boldsymbol{\zeta}^t\right) \leq \ln p(\boldsymbol{y}, \boldsymbol{\zeta}), \forall \boldsymbol{\zeta}$, $Q\left(\boldsymbol{\zeta}^t ; \boldsymbol{\zeta}^t\right)=\ln p\left(\boldsymbol{y}, \boldsymbol{\zeta}^t\right)$ and $\frac{\partial Q\left(\boldsymbol{\zeta}^t ; \boldsymbol{\zeta}^t\right)}{\partial \boldsymbol{\zeta}}=\frac{\partial \ln p\left(\boldsymbol{y}, \boldsymbol{\zeta}^t\right)}{\partial \boldsymbol{\zeta}}$. Therefore, we need to maximize $Q\left(\boldsymbol{\zeta} ; \boldsymbol{\zeta}^t\right)$ to update the next iterate $\boldsymbol{\zeta}^{t+1}$. Due to the non-convex nature of $Q\left(\boldsymbol{\zeta}; \boldsymbol{\zeta}^t\right)$, it is challenging to find the global optimal solution for the maximization problem. In this case, through the gradient ascent method, we can simply get the next update. 


Note that the differences between the SEA-LD-TurboCE algorithm and the previous turbo algorithm, such as \cite{ref10}, are as follows: 1) Instead of initializing with a uniform grid, the grid is initialized using prior information obtained from the long-timescale phase, leading to a more precise initial grid with fewer grid points. This approach enhances the estimation accuracy and reduces the complexity of the turbo algorithm, both in terms of the complexity of a single iteration and the total number of iterations needed. 2) The derivation of the algorithm under location domain modeling, which is a departure from the previous method that modeled in the angular domain. 3) The SEA-LD-TurboCE algorithm adopts non-orthogonal pilots instead of orthogonal pilots. The ability to distinguish the channels of different users under non-orthogonal pilots is facilitated by the location domain modeling and proper assignment of pilots.

\begin{algorithm}[H]
\caption{SEA-LD-TurboCE Algorithm.}
\label{alg:dbscan_clustering}
\begin{algorithmic}[1]
\STATE \textbf{Input}: $\boldsymbol{y}, \hat{\boldsymbol{\Xi}}, \boldsymbol{P}_{U\!E}^0$ maximum iteration numbers $I_{\text{in}}, I_{\text{out}},$ and  threshold $\epsilon$
\STATE \textbf{Initialization}:
\STATE Initialize the dynamic grid by the refined scatterer positions using $\boldsymbol{\gamma}^0= \hat{\boldsymbol{\Xi}}^{'}$.
\FOR{$t = 1, \ldots, I_{\text{out}}$}
    \STATE \textbf{E Step}:
    \STATE Initialize $i_{\text{in}} = 1, \boldsymbol{x}_A^{\text{pri}} = \mathbf{0}$ and $\boldsymbol{V}_A^{\text{pri}}$
    \WHILE{$i_{\text{in}} \leq I_{\text{in}}$}
        \STATE $i_{\text{in}} = i_{\text{in}} + 1$
        \STATE Update $\boldsymbol{x}_B^{\text{pri}} = \boldsymbol{x}_{A \rightarrow B}^{\text{ext}}$ and $\boldsymbol{V}_B^{\text{pri}} = \boldsymbol{V}_{A \rightarrow B}^{\text{ext}}$ using (33), (34).
        \STATE Perform message passing over the factor graph $\mathcal{G}_B$ using (41) - (46).
        \STATE Calculate $\hat{p}(x_q^k | \boldsymbol{y}, \boldsymbol{\zeta}^t), \hat{p}(s_q^k | \boldsymbol{y}, \boldsymbol{\zeta}^t), \forall q,k$ using (36), (37).
        \STATE Update $\boldsymbol{x}_A^{\text{pri}} = \boldsymbol{x}_{B \rightarrow A}^{\text{ext}}$ and $\boldsymbol{V}_A^{\text{pri}} = \boldsymbol{V}_{B \rightarrow A}^{\text{ext}}$, using (38), (39).
    \ENDWHILE
    \STATE \textbf{M Step}:
    \STATE Calculate the gradient $\frac{\partial Q(\boldsymbol{\zeta}^t;\boldsymbol{\zeta}^t)}{\partial \boldsymbol{\zeta}}$ using the $\boldsymbol{x}_B^{\text{post}}$ and $\boldsymbol{V}_B^{\text{post}}$ from the E step.
    \STATE Obtain $\boldsymbol{\zeta}^{t+1}$ using the gradient ascent.
    \IF{$\left\|\boldsymbol{\zeta}^{t+1} - \boldsymbol{\zeta}^t\right\| \leq \epsilon$}
        \STATE \textbf{break}
    \ENDIF
\ENDFOR
\STATE Output $\boldsymbol{\hat{\zeta}}, \boldsymbol{\hat{x}} = \boldsymbol{x}_B^{\text{post}}$ and $\hat{p}(s_q^k | \boldsymbol{y}, \boldsymbol{\hat{\zeta}}), \forall k$.
\end{algorithmic}
\end{algorithm}
\section{Numerical Results}\label{sec:simulation_results}
In this section, we provide numerical results to evaluate the channel estimation performance of the proposed scheme. The MIMO-OFDM system is equipped with a carrier frequency of $3.5$ GHz, the number of OFDM subcarriers is $P = 192$, and the subcarrier interval is $f_0 = 30$ kHz. By aggregating every 12 subcarriers into a single Resource Block (RB), the total number of RBs is $16$. The BS is equipped $N = 64$ ($N_x = N_y = 8$) antennas and is  at coordinates $[-50,0,25] \text{m}$. The number of users is $K=200$ and each user is uniformly distributed within a square area defined by the Cartesian coordinates \( x \in [50, 250] \, \text{m} \) and \( y \in [-100, 100] \, \text{m} \), with \( Z = 1.5 \, \text{m} \) representing the height above the ground.  We assume that the prior information of user locations is subject to errors that follow a Gaussian distribution and can be expressed as $\epsilon \sim \mathcal{N}(\mu, \sigma_{\text{U\!E}}^2)$, where $\sigma_{\text{U\!E}}$ is set as $5 m$ if not specified. Besides, the channel is generated according to the UMA model in the Quadriga platform model defined by 3GPP R16 specification~\cite{ref25} with $L=6$, and the performance results are obtained by averaging over 100 realizations. We choose the normalized mean square error (NMSE) as the performance metric to evaluate the extrapolation performance of various algorithms, which is defined as:
$\text{NMSE} = \frac{\| \hat{H} - H \|^2}{\| H\|^2}$. For comparison, we consider the following four benchmark schemes:
\begin{itemize}
  \item \textbf{MUSIC+LS:} The TST-MUSIC algorithm is used to estimate the positions of scatterers, and the complex gain is estimated by the LS algorithm.
 \item \textbf{Proposed scheme with MU and non-orthogonal pilots (MU-NP):} The MU channel estimation and localization are performed jointly using the proposed SEA-LD-TurboCE algorithm with the grid is initialized by TST-MUSIC and DBSCAN based scatterer association. Besides, the graph-coloring-based user grouping algorithm is performed for spatial pilot reuse.
   \item \textbf{Proposed scheme with MU and orthogonal pilots (MU-OP):} This scheme is the same as the proposed MU-NP scheme except that the different users employ orthogonal pilots thus there is no intra-group interference. 
  \item \textbf{Proposed scheme with single-user and orthogonal pilots (SU-OP):} The different users separately perform channel estimation and localization using the proposed SEA-LD-TurboCE algorithm (i.e., assuming independent sparse channel priors for different users). Besides, different users employ orthogonal pilots, and the grid is initialized by the TST-MUSIC algorithm.
 
  \item \textbf{Genie-aided LMMSE:} The Genie-aided LMMSE leverages perfect statistical knowledge of the channel to filter the received signals in both the time and spatial domains, and can be considered as an upper bound for the performance of channel estimation.
\end{itemize}

    \subsection{MU Scatterer Association Performance}
    In Fig. 5, we show the effectiveness of the DBSCAN-based MU scatterer association algorithm. In Fig. 5, clusters of different colors represent associated shared scatterers, while points identified as noise by DBSCAN are shown in gray. As can be seen, the coarsely estimated scatterer positions form a cluster near the true values, and the refined estimated positions, obtained by associating and updating the scatterers, are even closer to the true values when SNR=$0~$dB. Furthermore, the scatterer association algorithm can identify the global scatterers in the environment by combining the estimated results of MUs' scatterers, reducing the probability of missed detection as shown in Fig.9.

       \begin{figure}[H]
      \centering
      \includegraphics[width=4.5in]{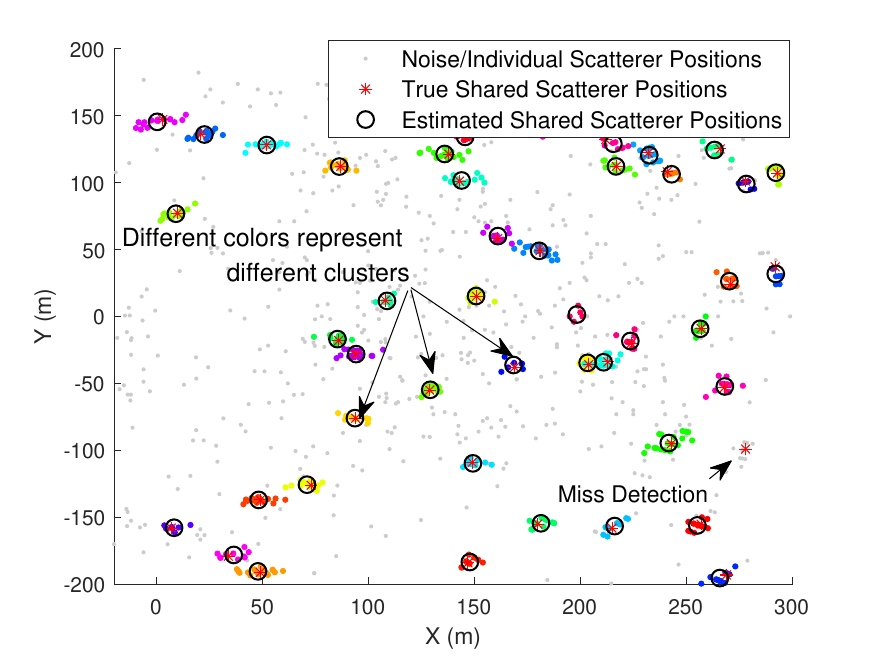}
      \caption{The performance of DBSCAN-based scatterer association algorithm (SNR= $0$ dB).}
      \label{figure_6}
    \end{figure}
    
\subsection{User grouping Performance}
The CDF of intra-group minimum distance is shown in Fig. 6, where the intra-group minimum distance is defined as the minimum distance defined in (11) between the two users in one group $\min_{k_1, k_2 \in \mathcal{G}_g} d_{k_1, k_2}$. Fig. 6 illustrates the effectiveness of the graph-coloring-based user grouping algorithm under true scatterer positions and estimated scatterer positions. It can be observed that the user grouping algorithm ensures the minimum distance between the true positions of scatterers within a group to reduce intra-group interference. Moreover, for the estimated positions of scatterers, it can be seen that approximately $10\%$ of users within the same group are at a distance of less than $50 m$ from one another, which can lead to intra-group interference. This is mainly caused by the inaccuracy in the estimation of scatterer positions and cannot be avoided through user grouping algorithms.


For the convenience of describing the degree of non-orthogonal pilot reuse, we define the pilot reduction factor \( \kappa = \frac{K}{G} \), which is the ratio of the number of users and the number of groups. The larger the value of \( \kappa \), the greater the multiple of spatial multiplexing pilots, and \( \kappa = 1 \) indicates the orthogonal pilot scheme. In Fig.7, the performance of the MU-NP scheme can be seen to decline with the increase of \( \kappa \) with different SNRs. This is because more users that reuse the same pilot will lead to increased interference among users within each group, thereby causing the NMSE to rise. Furthermore, by comparing the proposed MU-NP scheme with the SU-OP scheme, it can be observed that the SNR within the range of $-10~$dB to $10~$dB, there is approximately a $2.5$ to $4$ times increase in pilot reuse compared to the SU scheme. And as the SNR increases, the gain of MU-NP over SU-OP becomes smaller. The reason is that higher SNR makes the system more susceptible to interference within the group, and the partially non-orthogonal paths between users can significantly degrade channel estimation performance.

       \begin{figure}[H]
      \centering
      \includegraphics[width=4.5in]{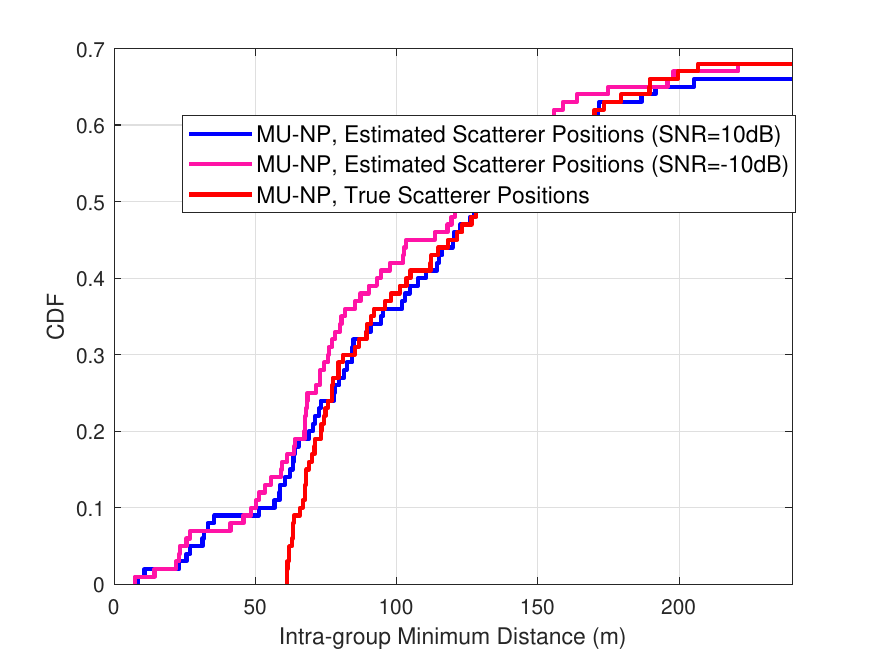}
      \caption{ The user grouping performance of the proposed algorithm.}
      \label{figure_6}
    \end{figure}
       \begin{figure}[H]
      \centering
      \includegraphics[width=4.5in]{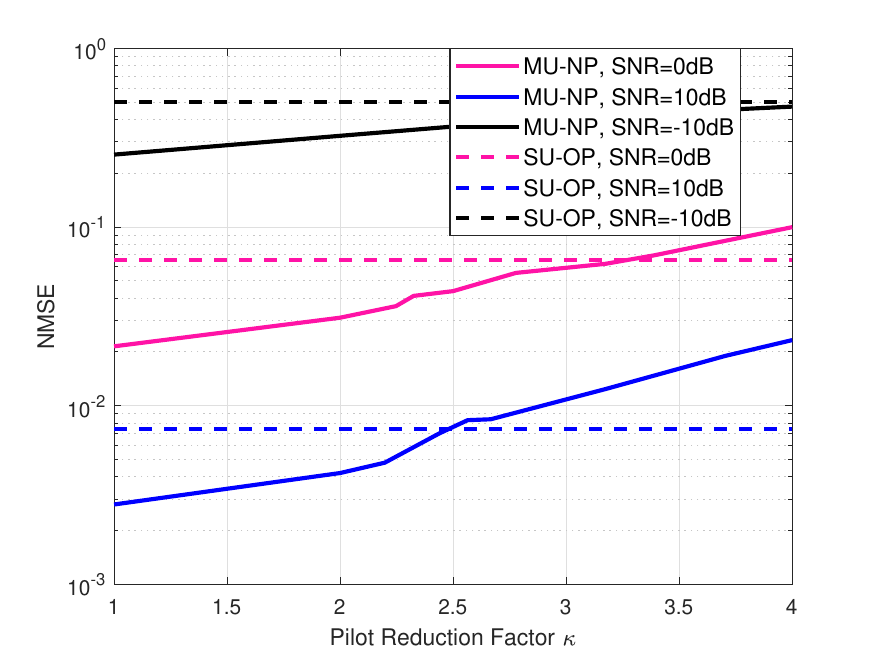}
      \caption{The NMSE performance versus pilot reduction factor.}
      \label{figure_6}
    \end{figure}

          \subsection{Channel Estimation Performance}
  In Fig. 8, we compare the NMSE performance of channel estimation versus SNR. The performance of all schemes improves as the SNR increases, but the performance of the MUSIC+LS scheme gradually levels off. This is because the MUSIC+LS scheme is unable to fully utilize the sparse information of the spatial domain channels. In addition, the MU-OP scheme achieves a significant performance gain over the SU-OP scheme and approaches the performance of the genie-aided scheme. This indicates that the SEA-LD-TurboCE algorithm can efficiently exploit the 3-D MU joint sparsity of the location domain channels. Furthermore, the MU-NP scheme achieves a relatively small performance loss compared to the MU-OP scheme, but the pilot resources can be reused twice through spatial division, demonstrating the effectiveness of the user grouping algorithm.
       \begin{figure}[H]
      \centering
      \includegraphics[width=4.5in]{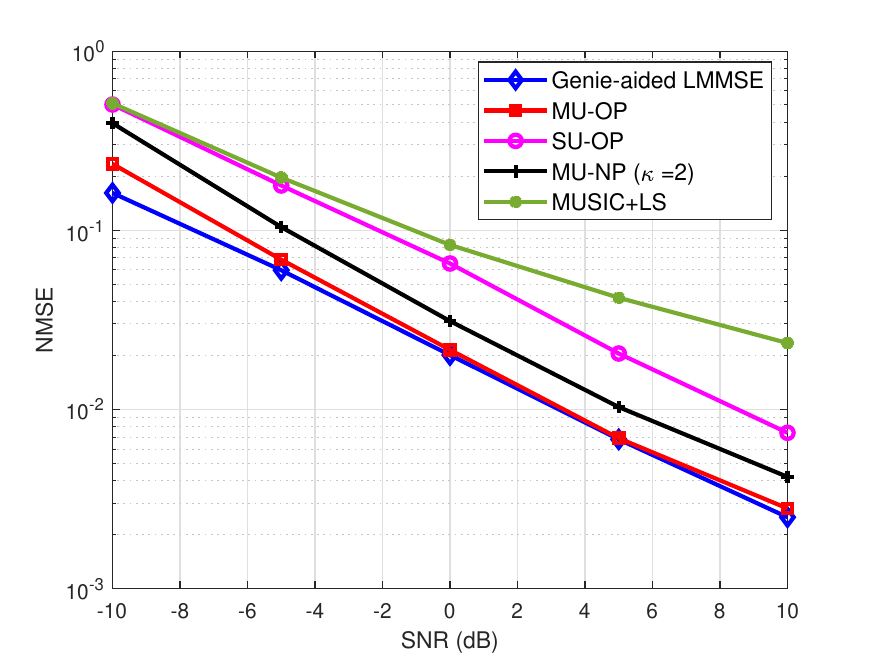}
      \caption{The NMSE performance of channel estimation versus SNR.}
      \label{figure_6}
    \end{figure}
       \subsection{Localization Performance}
   In Fig. 9 and Fig. 10, we compare the RMSE and miss-detection performance of scatterer localization versus SNR. It can be observed that the MU-NP scheme is capable of fully exploiting the overlapping characteristics of scatterers among MUs, thereby enhancing the precision of scatterer position estimation and reducing the miss detection probability. Moreover, when there is an error of $\sigma_{\text{U\!E}}=20~m$ in user location, the MU-NP scheme has a smaller loss in RMSE compared to the MUSIC+LS scheme, and at the same time, the false alarm probability is always less than that of the MUSIC+LS scheme. This is because the MU-NP scheme can utilize the overlapping information of scatterers among MUs, and thus, by fusing the position samples of multiple scatterers, it reduces the impact of user location errors.

\begin{figure}[H]
\centering    
\subfigure[RMSE performance] 
{
	\begin{minipage}{7cm}
	\centering          
	\includegraphics[width=3in]{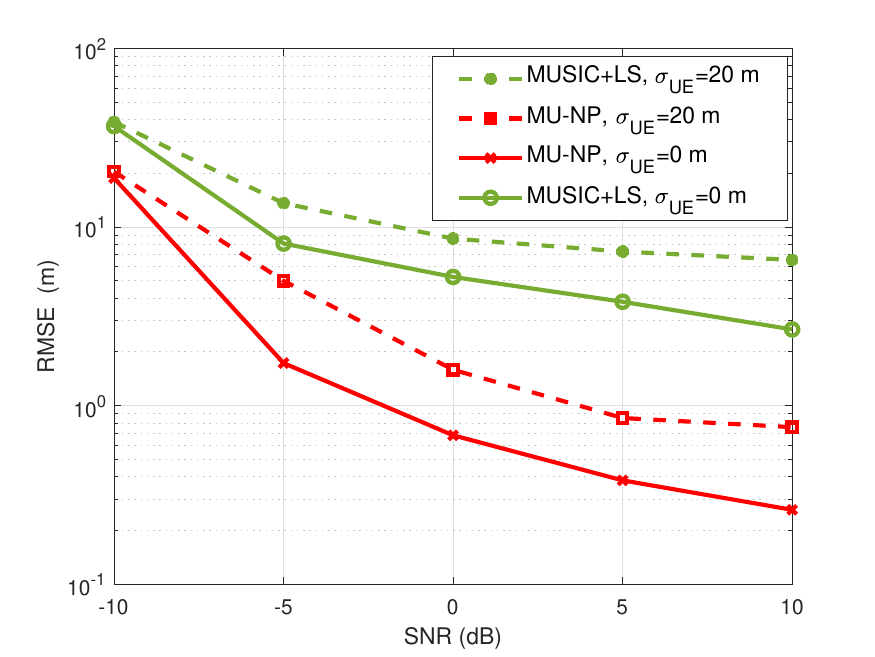}   
	\end{minipage}
}	
\subfigure[Miss detection performance] 
{
	\begin{minipage}{7cm}
	\centering      
	\includegraphics[width=3in]{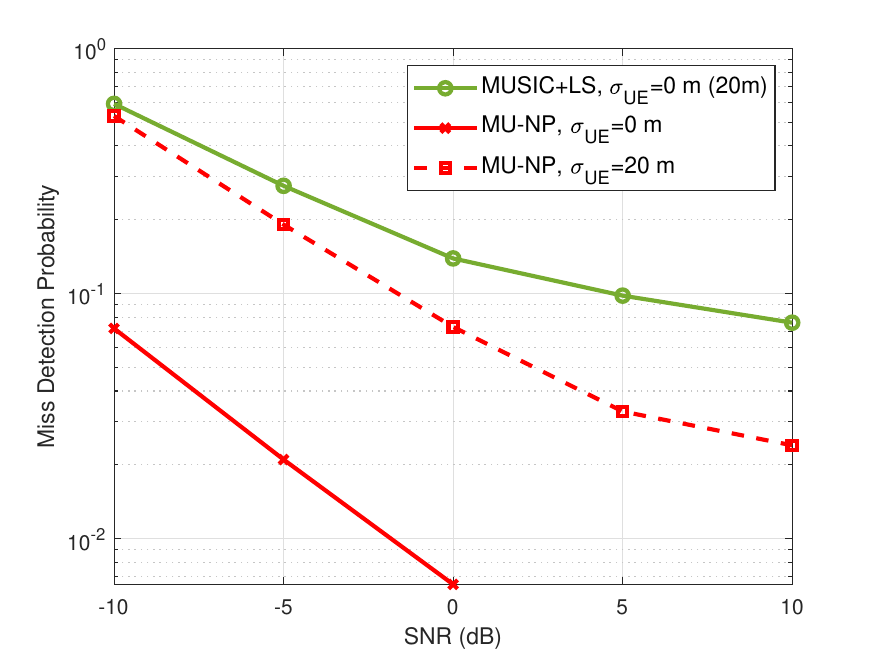}  
	\end{minipage}
}
\caption{The RMSE and miss detection performance of scatterer localization versus SNR.} %
\label{fig:1}  
\end{figure}

\section{Conclusion}\label{sec:conclusion}
In this paper, we propose a two-timescale approach for joint MU uplink channel estimation and localization in MIMO-OFDM systems, which fully captures the spatial characteristics of MUs. A location domain sparse representation of MU channels is introduced to fully utilize the overlapping and orthogonal channel characteristics of different MUs. In the long timescale phase, by merging the shared scatterer positions of MUs, we obtain the refined scatterer positions. Based on this information, we proposed a graph-coloring based user grouping algorithm, thus enabling spatial division multiplexing of pilots. In the short timescale phase, a low-complexity SEA-LD-TurboCE algorithm is introduced to facilitate high-precision joint MU channel estimation and localization under spatially reused pilots.  Simulation results verify the superior channel estimation and localization performance of our proposed scheme over the baselines and the proposed MU non-orthogonal pilots scheme can achieve more than double the pilot expansion gain compared to the SU scheme.

\appendix
\subsection{Message Update Equations for Module B of SEA-LD-TurboCE}	
1) Message Passing Over the Path $x_q^k \rightarrow f_q^k \rightarrow s_q^k \rightarrow \eta_q^k \rightarrow s_q \text { : }$
The message from factor node to variable $\eta_q^k$ node $s_q $ is
\begin{equation}
\begin{aligned}
\nu_{\eta_q^k \rightarrow s_q}\left(s_q\right) &=\sum_{\boldsymbol{s}^k} \eta_q^k\left(s_q^k, s_q\right) \times \nu_{s_q^k \rightarrow \eta_q^k}\left(s_q^k\right) \\
&=\pi_{s, q}^{k, i n} \delta\left(s_q-1\right)+\left(1-\pi_{s, q}^{k, i n}\right) \delta\left(s_q\right)
\end{aligned}    
\end{equation}
where $\pi_{s^k, q}^{k, i n}=\left(1+\frac{1-\pi_{s^k, q}^{i n}}{1+2 \pi_{s^k, q}^{i n}-\pi_{s^k, q}^{i n}-\rho^k}\right)^{-1}$, and $\pi_{s^k, q}^{i n}=\left(1+\frac{\mathcal{C N}\left(0 ; x_{B, q}^{k, p r i}, v_{B, q}^{k, p r i}\right)}{\mathcal{C N}\left(0 ; x_{B, q}^{k, p r i}, v_{B, q}^{k, p r i}+\left(\sigma_q^k\right)^2\right)}\right)^{-1}$.

2) Message Passing Over the Path $x_0^k \rightarrow f_0^k \rightarrow s_0^k \rightarrow h_0^k \text { : }$
The message from factor node to variable $s_0^k$ node $h_0^k $ is
\begin{equation}
\begin{aligned}
\mathcal{V}_{s_0^k \rightarrow h_0^k}\left(s_0^k\right)=\mathcal{V}_{f_0^k \rightarrow s_0^k}\left(s_0^k\right) & \propto \int \mathcal{V}_{x_0^k \rightarrow f_0^k}\left(x_0^k\right) \times f_0^k\left(x_0^k, s_0^k\right) d x_0^k \\
&=\pi_{s_0^k}^{i n} \delta\left(s_0^k-1\right)+\left(1-\pi_{s_0^k}^{i n}\right) \delta\left(s_0^k\right)
\end{aligned}\end{equation}
where$\pi_{s_0^k}^{i n}=\left(1+\frac{\mathcal{C N}\left(0 ; x_{B, 0}^{r, p r i}, v_{B, 0}^{r, p r i}\right)}{\mathcal{C N}\left(0 ; x_{B, 0}^{r, p r i}, v_{B, 0}^{r, p r i}+\left(\sigma_0^k\right)^2\right)}\right)^{-1}$.

3) Message Passing Over the Path $s_q \rightarrow h_q \text { : }$
The message from factor node to variable $s_q$ node $ h_q$ is
\begin{equation}\begin{aligned}
\mathcal{V}_{s_q \rightarrow h_q}\left(s_q\right) & \propto  \prod \limits_{k \in \mathcal{K}_q} \mathcal{V}_{\eta_{q \rightarrow s_q}^k}\left(s_q\right) \\
&=\pi_q^{i n} \delta\left(s_q-1\right)+\left(1-\pi_q^{i n}\right) \delta\left(s_q\right),
\end{aligned}\end{equation}
where $\pi_q^{i n}=\frac{\prod \limits_{k \in \mathcal{K}_q} \pi_{s, q}^{k, i n} }{\prod \limits_{k \in \mathcal{K}_q} \pi_{s, q}^{k, i n} +\prod \limits_{k \in \mathcal{K}_q} \left(1-\pi_{s, q}^{k, i n}\right)}$, and $\mathcal{K}_q $ denotes the set of factor nodes that are neighbors of $s_q$ except $h_q$.

4) Message Passing Over the Path $h_q \rightarrow s_q \text { : }$
The message from factor node to variable $h_q$ node $ s_q$ is
\begin{equation}
\begin{aligned}
\mathcal{V}_{h_q \rightarrow s_q} &\left(s_q\right) \propto h_q\left(s_q\right) \\
&=\lambda \delta\left(s_q-1\right)+(1-\lambda) \delta\left(s_q\right).
\end{aligned}
\end{equation}

5) Message Passing Over the Path $s_q \rightarrow \eta_q^k \rightarrow s_q^k\rightarrow f_q^k \rightarrow x_q^k \text { : }$
The message from factor node to variable $ f_q^k$ node $ x_q^k$ is
\begin{equation}
\begin{aligned}
\mathcal{V}_{f_q^k \rightarrow x_q^k}\left(x_q^k\right) & \propto \sum_{s_q^k} \mathcal{V}_{s_q^k \rightarrow f_q^k}\left(s_q^k\right) \times f_q^k\left(x_q^k, s_q^k\right) \\
&=\pi_{s^k, q}^{o u t} \mathcal{C N}\left(x_q^k ; 0,\left(\sigma_q^k\right)^2\right)+\left(1-\pi_{s^k, q}^{o u t}\right) \delta\left(x_q^k\right).
\end{aligned}
\end{equation}
where $\pi_{s^k, q}^{o u t}=\rho^k \pi_{s, q}^{k, o u t}$, and $\pi_{s, q}^{k, o u t}=\frac{\lambda \prod \limits_{l \in \mathcal{K}_q\backslash k} \pi_{s, q}^{l, i n}}{\lambda \prod \limits_{l \in \mathcal{K}_q \backslash k} \pi_{s, q}^{l, i n}+(1-\lambda)\prod \limits_{l \in \mathcal{K}_q\backslash k} \left(1-\pi_{s, q}^{l, i n}\right)}$.

6) Message Passing Over the Path $h_0^k \rightarrow s_0^k \rightarrow f_0^k \rightarrow x_0^k \text { : }$
The message from factor node to variable $ f_0^k $ node $  x_0^k$ is

\begin{equation}
\begin{aligned}
\mathcal{V}_{f_0^k \rightarrow x_0^k}\left(x_0^k\right) & \propto \sum_{s_0^k} \mathcal{V}_{s_0^k \rightarrow f_0^k}\left(s_0^k\right) \times f_0^k\left(x_0^k, s_0^k\right) \\
&=\rho_0^k \mathcal{C N}\left(x_0^k ; 0,\left(\sigma_0^k\right)^2\right)+\left(1-\rho_0^k\right) \delta\left(x_0^k\right)
\end{aligned}
\end{equation}

\end{document}